\documentclass[twocolumn,reprint,amsmath,amssymb,footinbib,superscriptaddress,floatfix,aps,longbibliography,prx]{revtex4-2}
\pdfoutput=1

\usepackage{textcomp}
\usepackage[utf8]{inputenc}
\usepackage[sort&compress]{natbib}
\usepackage[dvipsnames]{xcolor}
\usepackage{graphicx}
\usepackage{amssymb}
\usepackage{epsfig}
\usepackage{epstopdf}
\usepackage{amsfonts}
\usepackage{xcolor}
\usepackage{amstext,amsmath,amssymb}
\usepackage{soul}
\usepackage{rotating}
\usepackage{dcolumn}
\usepackage{array,multirow}
\usepackage[english]{babel}
\usepackage[export]{adjustbox}
\usepackage{braket}
\usepackage[caption=false,subrefformat=parens,labelformat=parens]{subfig}
\usepackage{xr}
\usepackage{ifpdf}
\usepackage{tikz}
\usetikzlibrary{arrows}
\usepackage{amsmath,amssymb,amsfonts}
\usepackage{xcolor,graphicx}
\usepackage[bookmarks=false,linkcolor=cyan,urlcolor=blue,colorlinks,citecolor=cyan]{hyperref}

\begin{document}

\title{Robust gates with spin-locked superconducting qubits}

\author{Ido Zuk}
\affiliation{AWS Center for Quantum Computing, Pasadena, California 91125, USA}
\affiliation{Racah Institute of Physics, The Hebrew University of Jerusalem, Jerusalem 91904, Givat Ram, Israel}

\author{Daniel Cohen}
\affiliation{AWS Center for Quantum Computing, Pasadena, California 91125, USA}
\affiliation{Racah Institute of Physics, The Hebrew University of Jerusalem, Jerusalem 91904, Givat Ram, Israel}

\author{Alexey V.
Gorshkov}
\affiliation{AWS Center for Quantum Computing, Pasadena, California 91125, USA}

\author{Alex Retzker}
\affiliation{AWS Center for Quantum Computing, Pasadena, California 91125, USA}
\affiliation{Racah Institute of Physics, The Hebrew University of Jerusalem, Jerusalem 91904, Givat Ram, Israel}

\begin{abstract}
Dynamical decoupling is effective in reducing gate errors in most quantum computation platforms and is therefore projected to play an essential role in future fault-tolerant constructions. In superconducting circuits, however, it has proven difficult to utilize the benefits of dynamical decoupling. In this work, we present a theoretical proposal that incorporates a continuous version of dynamical decoupling, namely spin locking, with a coupler-based CZ gate for transmons and provide analytical and numerical results that demonstrate its effectiveness.
\end{abstract}
\maketitle

\section{Introduction}
Qubit coherence time ($T_2$) is a major limitation on gate fidelities and a bottleneck for realizing a fault-tolerant quantum computer. Although the theoretical limit is $T_2\leq2T_1$ (where $T_1$ is the relaxation time), in most superconducting circuits to date, this bound is not saturated, which leaves substantial room for improvement of dephasing time, $T_\phi$, which relates to $T_2$ via $\frac{1}{T_2}=\frac{1}{T_\phi}+\frac{1}{2T_1}$. The limiting factors for $T_\phi$ in superconducting circuits \cite{Krantz_2019_revised} are flux fluctuations (in flux tunable qubits) \cite{PhysRevLett.97.167001}, charge coupling to two-level systems (TLSs) \cite{osman2023mitigation_revised2,bilmes2022probing_revised2,lisenfeld2019electric_revised3,phillips1987two_revised1}, dephasing due to the readout resonator \cite{PhysRevA.74.042318,PhysRevA.79.013819,PhysRevLett.120.260504_revised}, and ZZ cross-talk with other qubits.

Pulsed dynamical decoupling 
is a well-known method to enhance coherence time \cite{RevModPhys.88.041001}, utilized in various platforms
\cite{PhysRevB.85.155204,Biercuk2009,5500699,PhysRevA.82.042306, Peng_2011_revised, Jenista2009_rev2}. Its major use with superconducting qubits attempted to improve the idling stage \cite{Souza2021_revised, PhysRevLett.121.220502, Jurcevic_2021_revised}. 
Dynamical decoupling was also used to reduce unitary and spectator errors during cross-resonance gates with superconducting qubits \cite{PhysRevApplied.18.024068_revised3,PhysRevLett.117.210505, PhysRevLett.119.180501, PRXQuantum.1.020318}. However, the embedding of pulses into this gate affects the gate's speed and therefore makes the gate more susceptible to $T_1$ errors. Additionally, in current implementations of this gate, the qubits are required to be in the straddling regime \cite{Krantz_2019_revised}, which complicates frequency allocation in the device and introduces strong unwanted $ZZ$ interaction. On the other hand, the CZ gate in superconducting qubits is fast- but typically utilizes an additional state outside the computational subspace \cite{PhysRevX.11.021058,PhysRevA.90.022307,10.1126/sciadv.aao3603_revised,PhysRevLett.123.120502}. Thus- dynamical decoupling during such a gate would need to include the extra level, which is a major complication. Alternatively, one could use a gate from the iSWAP family \cite{PhysRevLett.125.120504}, which does not leave the computational subspace and therefore could be integrated with dynamical decoupling pulses. However, this type of gate also requires working inside the straddling regime and suffers from the same difficulties explained above.
  
Another version of dynamical decoupling is spin locking, where the qubit is driven continuously \cite{goldman1970spin,PhysRevLett.2.301,PhysRevA.77.052334,cai2012robust_revised, PhysRevLett.109.070502, Barfuss2018_1}. In order to gain (in terms of fidelity) from a spin-locked two-qubit gate between two qubits with coupling $g$ in the presence of phase noise with strength $\sigma_f$, the drive Rabi frequency $\Omega$ must satisfy $\Omega\gg g,\sigma_f$. While in trapped ions a two-qubit gate that fulfills these requirements was proposed and experimentally demonstrated \cite{PhysRevA.85.040302, PhysRevLett.110.263002,Timoney2011_1, PhysRevLett.117.220501}, in superconducting transmon qubits such a gate has remained elusive. The main difficulty in driving a transmon qubit with a large $\Omega$, which is comparable to the anharmonicity, is the increased leakage rate outside of the computational basis. 
 
Dressing superconducting qubits has been previously demonstrated experimentally \cite{PhysRevLett.98.257003}. It was used for noise spectroscopy \cite{Yan2013_revised,PRXQuantum.1.010305,PhysRevLett.120.260504_revised,Sung2021_revised}, was proposed for overcoming photon loss to TLSs \cite{PhysRevA.105.062605}, was analyzed using flux drive of the fluxonuim qubit \cite{PhysRevApplied.15.034065}, and was used to realize a two-qubit gate mediated by a bus resonator \cite{PRXQuantum.3.040322, PhysRevLett.121.130501}. However, to the best of our knowledge, it has never been shown how to use spin locking with a strong drive ($\Omega\gg g,\sigma_f$) to protect superconducting transmon qubits from dephasing during all relevant operations and gates.

In this paper, we propose a method to incorporate a strong spin locking drive into quantum computing architectures based on superconducting transmon qubits, which in turn enables high-fidelity one-qubit and two-qubit gates that are protected from dephasing. Our proposed method is designed to be incorporated into the coupler-based bus-below-qubit architecture \cite{PhysRevLett.127.080505}, and does not require working in the straddling regime. Moreover, the proposed method is \textit{modular}, which means that using a chip that enables the bus-below-qubit architecture, the experimentalist could select the qubits that need protection, and spin lock only them. Our method does not require the utilization of fixed-frequency transmons as in \cite{PhysRevLett.127.080505}, as it is not limited by the short dephasing time of tunable transmons.

The paper is organized as follows. First, in Sec.~\ref{section: Spin locking of a transmon}, we define the criteria that the drive needs to satisfy to reduce sensitivity to phase noise from arbitrary sources. Then, in Sec.~\ref{section: Coherence time of the driven qubit}, we demonstrate the robustness of the spin-locked qubit to phase noise in the idling regime, and discuss the limitations of our method due to noise in the drive amplitude. In Sec.~\ref{section: Full model - cosine potential}, we explain how we used the full cosine model of the transmon for gate simulations. We continue in Sec.~\ref{section: one-qubit gate} with a recipe for performing one-qubit gates in the dressed basis using the derivative-removal-by-adiabatic-gate (DRAG) \cite{PhysRevLett.103.110501, PhysRevA.83.012308}. Then, in Sec.~\ref{section: two-qubit gate}, we discuss the two-qubit gate, and show how to implement the adiabatic ZZ gate in the dressed basis using the bus-below-qubit architecture \cite{PhysRevLett.127.080505}. Then, in Sec.~\ref{section: Phase noise and data vs.~ancilla errors} we show that, in a surface code \cite{doi:10.1063/1.1499754_revised2, PhysRevA.86.032324} context where ancilla qubits are used to extract syndrome measurements on data qubits, spin-locking only the data qubits is enough, as the propagation of phase noise from the ancilla to the data is suppressed. This reduces the experimental complexity of driving both data and ancilla qubits, with only a small change in the code's threshold \cite{doi:10.1126/science.279.5349.342_revised, doi:10.1137/S0097539799359385_revised, KITAEV20032_revised}, and also avoids the complication of frequent measurements of spin-locked qubits. Lastly, in Sec.~\ref{section: State preparation and measurement in the dressed basis}, we show how to prepare and measure the dressed states. In Sec.~\ref{section: Leakage detection with the SL qubit}, we show how to detect leakage from the computational dressed states. In Sec.\ \ref{sec:disc}, we present the discussion. Appendices present details omitted from the main text. 
\begin{figure}[b]
\centering
\includegraphics[width=8.6cm]{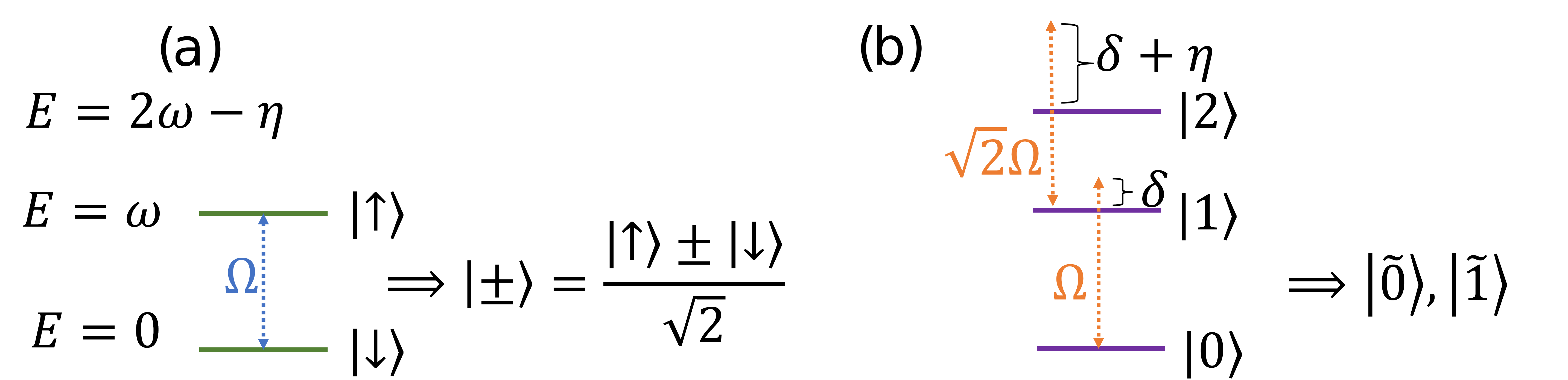}
\caption{Energy diagrams showing spin locking (a) of a two-level system and (b) of a transmon, using a continuous drive with Rabi frequency $\Omega$. The two-level system is driven on-resonance to create the protected dressed states $\ket{\pm}$. The transmon qubit must be driven off-resonantly, with detuning $\delta>0$, to create the protected dressed states $\ket{\tilde{0}}, \ket{\tilde{1}}$. $\omega$ is the energy difference between the lowest two levels of both systems, and $\eta$ is the anharmonicity of the transmon.}
\label{fig:energy_levels}
\end{figure}

\begin{figure}[htp]
\centering
\includegraphics[width=8.6cm]{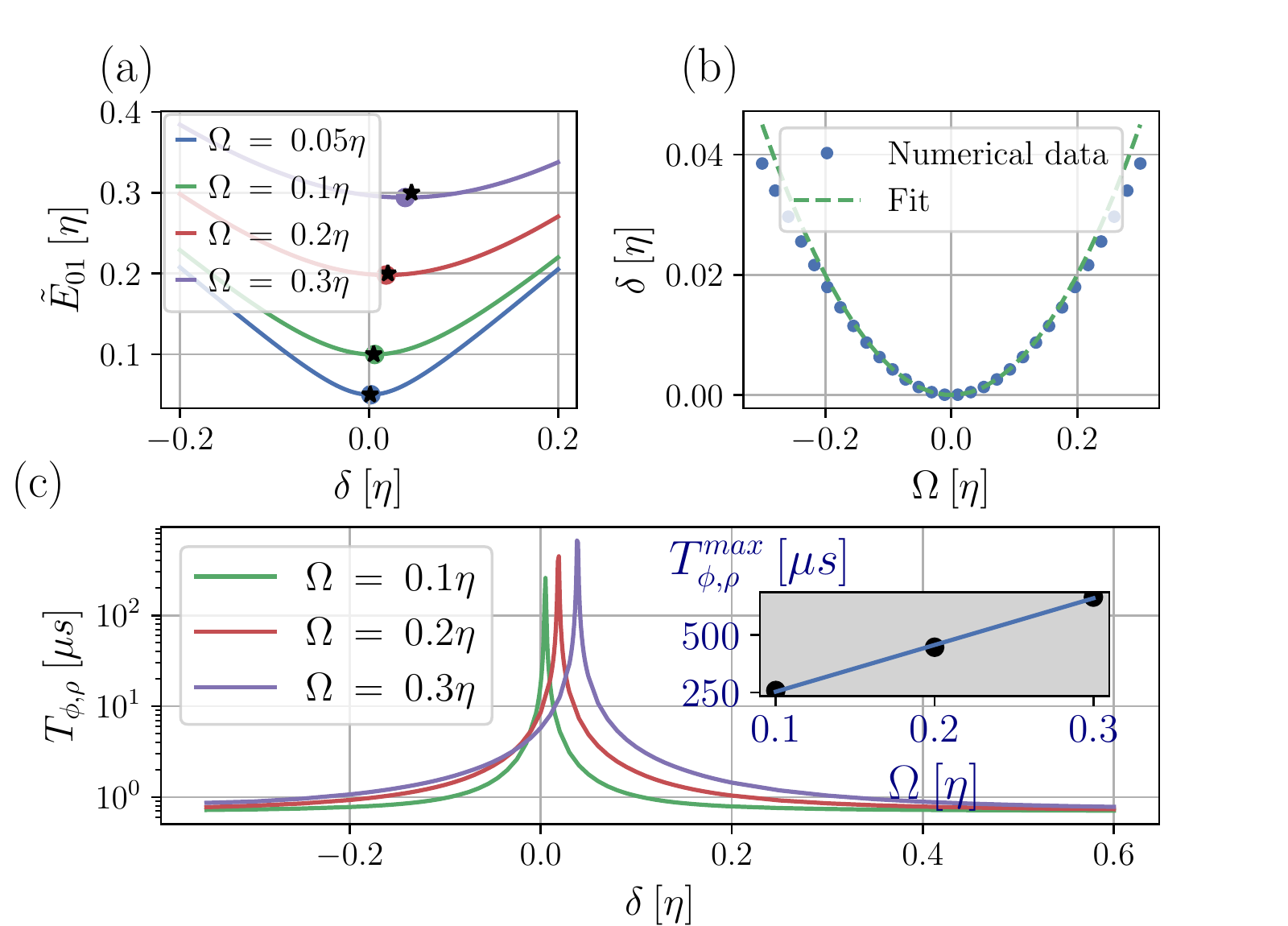}
\caption {(a) $\tilde{E}_{01}$ as a function of $\delta$, for several Rabi frequencies $\Omega$. The numerical minima are marked with circles. The stars are the analytical predictions (correct to order $\epsilon^2$) for the locations of the minima: $\delta = \Omega^2/(2 \eta)$ and $\tilde{E}_{01} = \Omega$. (b) $\delta$ for the clock condition for different $\Omega$. The dashed line is a fit $\frac{\delta}{\eta}=m\left(\frac{\Omega}{\eta}\right)^{2}$ to the points with $|\Omega|\leq0.03\eta$, which yields $m=\frac{1}{2}\pm0.002$. (c) $T_{\phi,\rho}$ as function of $\delta$, as obtained for $\frac{1}{f}$ noise. The maximal $T_{\phi,\rho}$ scales linearly with $\Omega$ (see inset, which shows  $T_{\phi,\rho}^{max}=Max_{\delta}\{T_{\phi,\rho}\}$ as a function of $\Omega$).  
The numerical results were obtained
in (a)(b) with 10 Fock states and in (c) with five Fock states. The precision in all of these results is better than 1\%.}

\label{fig:t_solution}
\end{figure}

\section{Spin locking of a transmon}
\label{section: Spin locking of a transmon}
The difficulty in using spin-locking with transmon qubits can be understood as follows. Suppose we realize the standard spin-locking scheme by continuously driving the transmon on resonance with the qubit frequency. For a two-level system, that would make the dressed basis resilient to phase noise. However, driving a transmon this way would also create populations in the higher Fock states \cite{Sung2021_revised,PhysRevLett.110.263601, PhysRevX.13.011008}, which would generally imply linear sensitivity to phase noise. While this could be addressed using extensive coherent control, such an approach would be experimentally challenging. Figure \ref{fig:energy_levels} shows the difference between standard spin locking for a two-level system and the transmon case.

Therefore, we start by deriving a more general condition for minimizing phase-noise sensitivity of spin-locked superconducting qubits. The Hamiltonian of a transmon driven with frequency $\omega_d$ and Rabi frequency $\Omega$, within the rotating-wave approximation ($\Omega\ll\omega_d$) and in the rotating frame, is
\begin{equation}
H =-\delta a^{\dagger}a-\frac{\eta}{2}a^{\dagger2}a^{2}+\frac{\Omega}{2}(a+a^{\dagger}).\label{eq:RWA_Hamiltonian}
\end{equation}
Here $a^\dagger$ is a bosonic creation operator, $\eta>0$ is the anharmonicity, and $\delta=\omega_d-\omega$, where $\omega$ is the qubit frequency. Since $\eta>0$, it is natural to order the dressed eigenenergies (eigenstates) of this Hamiltonian $\{\tilde{E}_i\}$ ($\{|\tilde{i}\rangle\}$) from the highest to the lowest. To have insensitivity to phase noise, we require $\frac{d\tilde{E}_{01}}{d\delta}=0$ for $\tilde{E}_{01}=\tilde{E}_{0}-\tilde{E}_{1}$. This translates to choosing $\delta$ satisfying
\begin{equation}
\langle\tilde{0}|a^{\dagger}a|\tilde{0}\rangle=\langle\tilde{1}|a^{\dagger}a|\tilde{1}\rangle.\label{eq:insensativity}
\end{equation}
We will refer to Eq.~(\ref{eq:insensativity})  as the \textit{clock condition} \cite{PhysRevLett.95.060502} and to $|\langle\tilde{0}|a^{\dagger}a|\tilde{0}\rangle-\langle\tilde{1}|a^{\dagger}a|\tilde{1}\rangle|$ as the \textit{sensitivity measure} (which is proportional to $|\frac{d\tilde{E}_{01}}{d\delta}|$).  Under the clock condition, the dressed qubit is insensitive to phase fluctuations to first order. We are interested in finding a function $\delta(\Omega)$ that, for any given $\Omega$, gives a detuning $\delta$ for which the clock condition holds. In the limit $\eta\rightarrow\infty$, we have $\delta(\Omega)=0$, and the dressed states are simply $(\ket{0} \pm \ket{1})/\sqrt{2}$. Therefore, as long as $\Omega\ll\eta$, we expect to find small  $\delta(\Omega)$. Since conjugating $H$ by a unitary $e^{-i\pi a^{\dagger}a}$ results in $\Omega\rightarrow-\Omega$, $\delta(\Omega)$ must be an even function. In particular, we find 
\begin{equation}
\frac{\delta(\Omega)}{\eta}=\frac{1}{2}\left(\frac{\Omega}{\eta}\right)^{2}+ \mathcal{O}\left(\left(\frac{\Omega}{\eta}\right)^4 \right). 
\label{eq:delta_up_to_four}
\end{equation}
The term $\frac{1}{2} (\frac{\Omega}{\eta})^{2}$ ensures that the $\ket{0}$-$\ket{1}$ transition is driven on resonance when the Stark shift $\frac{\Omega^2}{2\eta}$ of state $\ket{1}$ due to state $\ket{2}$  is taken into account. By diagonalizing Eq.~(\ref{eq:RWA_Hamiltonian}) under the condition in Eq.~(\ref{eq:delta_up_to_four}), to second order in $\epsilon = \Omega/\eta$, we find
\begin{align}
 & \tilde{E_{0}}=\frac{\Omega}{2}+ \mathcal{O}(\epsilon^3)  ,\\
 & \tilde{E_{1}}=-\frac{\Omega}{2}+ \mathcal{O}(\epsilon^3),\nonumber \\
 & \tilde{E_{2}}=\eta \left(-1-\frac{9}{8}\epsilon^{2}+ \mathcal{O}(\epsilon^3)\right),\nonumber \\
 & \tilde{E_{3}}=\eta \left(-3-\frac{37}{24}\epsilon^{2}+ \mathcal{O}(\epsilon^3)\right).\nonumber
\end{align}

In Appendix \ref{appendix:Eigenenergies and eigenstates of the SL qubit}, we show the eigenstates of Eq.~(\ref{eq:RWA_Hamiltonian}) under Eq.~(\ref{eq:delta_up_to_four}), and the creation operator in the dressed basis.
Figure \ref{fig:t_solution}(a) shows $\tilde{E}_{01}$ as function of $\delta$, for different $\Omega$, while $\delta = \Omega^2/(2 \eta)$ is marked with black stars. We see that, for $|\Omega|\lessapprox0.2 \eta$, $\delta = \Omega^2/(2 \eta)$ approximates well the location of the minimum, while for, larger $|\Omega|$, higher orders in Eq.~(\ref{eq:delta_up_to_four}) become important. Fig.~\ref{fig:t_solution}(b) shows the numerical values of $\delta$ that fulfill the clock condition for various $\Omega$, with good agreement with Eq.~(\ref{eq:delta_up_to_four}) for small enough Rabi frequency. 

We note that our method can also deal with phase noise originating from another system that is dispersively coupled to the qubit. We assume we can approximate the interaction between the spin-locked qubit and the second system as $\chi a^\dagger a b^\dagger b + \mathcal{O}(\epsilon)$, where $\chi$ is the interaction strength and $b$ is the second-system annihilation operator. Therefore, to zeroth order in $\epsilon$, the clock condition [Eq.~(\ref{eq:insensativity})] ensures protection from fluctuations in the second-system population. In Appendix \ref{appendix:Transverse coupling}, we give more details regarding the validity of this approximation. There are two prominent examples of this type of noise: dephasing due to the readout resonator \cite{PhysRevA.74.042318,PhysRevA.79.013819} and dephasing due to TLSs \cite{Krinner2022}. Additionally, photon loss due to a near-resonant TLS \cite{Burnett2019_revised, Carroll2022_revised,osman2023mitigation_revised2} could also be addressed by spin locking \cite{PhysRevA.105.062605}, and specifically with our method, by choosing $\Omega$ to avoid the resonance with the TLS.

\section{Coherence time of the driven qubit}
\label{section: Coherence time of the driven qubit}

In Fig.~\ref{fig:t_solution}(c), we study the dephasing time $T_{\phi,\rho}$ of the spin-locked qubit in the presence of $\frac{1}{f}$ noise \cite{RevModPhys.86.361}, i.e., we simulate the Hamiltonian in Eq.~(\ref{eq:RWA_Hamiltonian}) with a stochastic term $p(t)a^{\dagger}a$, where $p(t)$ is pink noise. We set the variance of the noise such that the dephasing time $T_\phi$ of the bare transmon is $T_\phi\approx 700$ ns (which is in the range of realistic values for a tunable transmon operating on the slope of the cosine potential \cite{Krantz_2019_revised}). We see that spin locking improves the dephasing time by a factor of 300--600. Moreover, we see that, for $\Omega/\eta = 0.2, 0.3$, it is important to set $\delta \neq 0$ to reach the best performances.
Let us analyze the curvature of the spin-locked qubit energy gap $\tilde{E}_{01}=\tilde{E}_0-\tilde{E}_1$ in the presence of phase noise. Consider varying $\delta$ [see Eq.~(\ref{eq:RWA_Hamiltonian})] away from the clock condition [Eq.~(\ref{eq:insensativity})]. By definition of the clock condition, the first derivative of $\tilde{E}_{01}$ vanishes:  $\frac{d\tilde{E}_{01}}{d\delta}=0$. In the limit of small $\Omega$, the second derivative can be found by analyzing only state $\ket{0}$ and the shifted state $\ket{1}$. In this basis, the Hamiltonian is $\frac{\Omega}{2}\sigma_x - \delta' \ket{1} \bra{1}$, where $\delta'$ is the deviation of $\delta$ from the clock condition. By moving to the $\sigma_x$ basis and treating $\delta'$ as a perturbation, we arrive at $\frac{d^2\tilde{E}_{01}}{d\delta'^2}=1/\Omega + \mathcal{O}(\Omega^0)$. Therefore, increasing $\Omega$ decreases the sensitivity to both frequency error and phase noise. Fig.~\ref{fig:t_solution}(a) shows that $\frac{d^2\tilde{E}_{01}}{d\delta^2}$ indeed decreases with increasing $\Omega$, and the inset in Fig.~\ref{fig:t_solution}(c) shows the resulting linear relation between the maximal $T_{\phi,\rho}$ and $\Omega$.

The bare dephasing time $T_\phi$ is inversely proportional to the noise standard deviation $\sigma_f$. As we have just derived, $\tilde E_{01}$ depends only quadratically---with coefficient $1/\Omega$---on frequency shifts away from the clock condition. The dressed qubit thus effectively feels reduced noise $\sim \sigma_f^2/\Omega$. Therefore, in the absence of noise in the drive amplitude, the dressed dephasing time obeys $T_{\phi,\rho}\propto \frac{\Omega}{\sigma_f^2}=(\Omega T_\phi)T_\phi$. Thus, we expect the improvement factor of 300--600 to increase even further with $T_\phi$ (see Sec.~\ref{subsec: limiting factors} for a discussion of noise in $\Omega$). In our example [Fig.~\ref{fig:t_solution}(c)], $T_\phi=700$ ns is near the minimum for known systems, and thus we expect this improvement factor to increase in relevant experimental systems (in the absence of noise on the drive amplitude). 

\subsection{Limiting factors for the coherence time of the driven qubit}
\label{subsec: limiting factors}
Spin locking reduces the noise effects as follows: The transition rate between the dressed states ($T_{1\rho}^{-1}$) is dominated by the power spectrum of the noise at the dressed states' gap frequency, $S\left(\Omega+\mathcal{O}((\frac{\Omega}{\eta})^3)\right)$, and the original $T_1^{-1}$. We expect $T_1^{-1}$ to dominate over $S\left(\Omega+\mathcal{O}((\frac{\Omega}{\eta})^3)\right)$ (due to small noise at high frequencies in superconducting systems \cite{PhysRevLett.130.220602, Yan2016_1}) and therefore determine $T_{1\rho}^{-1}$, which provides the spin-locked qubit robustness against phase noise and thus high degree of tunability as it relaxes the restriction of operating in the vicinity of the sweet spot. The dephasing rate $T_{\phi,\rho}^{-1}$ of the spin-locked qubit will be dominated by two factors: (1) drive-amplitude noise; and (2) frequency noise, most likely due to flux noise (for tunable transmons) and TLSs \cite{osman2023mitigation_revised2,bilmes2022probing_revised2,lisenfeld2019electric_revised3,phillips1987two_revised1} and the above-discussed second-order phase noise. Regarding (1), the fractional noise amplitude is usually of order $10^{-3}$ which, for $\frac{\Omega}{2\pi} = 50$ MHz, corresponds to $T_{\phi,\rho} \approx 125$ $\mu$s. Although this is a significant coherence limitation, due to its long correlation time, amplitude noise will only contribute to the infidelity at the level of $\left( \frac{T_g}{125 \mu\textrm{s}} \right)^2$ (where $T_g$ is the time of the gate), which is below the amplitude-damping contribution. Moreover, the long correlation time of this noise allows for the utilization of efficient decoupling techniques \cite{PhysRevLett.2.301,cai2012robust_revised,Cohen_2016}. With extra effort, the fractional noise could be reduced to the $10^{-4}$ level \cite{Hacohen-Gourgy2016,10.1126/science.abc5186_revised, PhysRevX.13.011008}, which will increase the dephasing time to $T_{\phi,\rho} \approx 1250$ $\mu$s.

\section{Full model - cosine potential}
\label{section: Full model - cosine potential}

\begin{figure}[t]
\includegraphics[width=8.6cm]{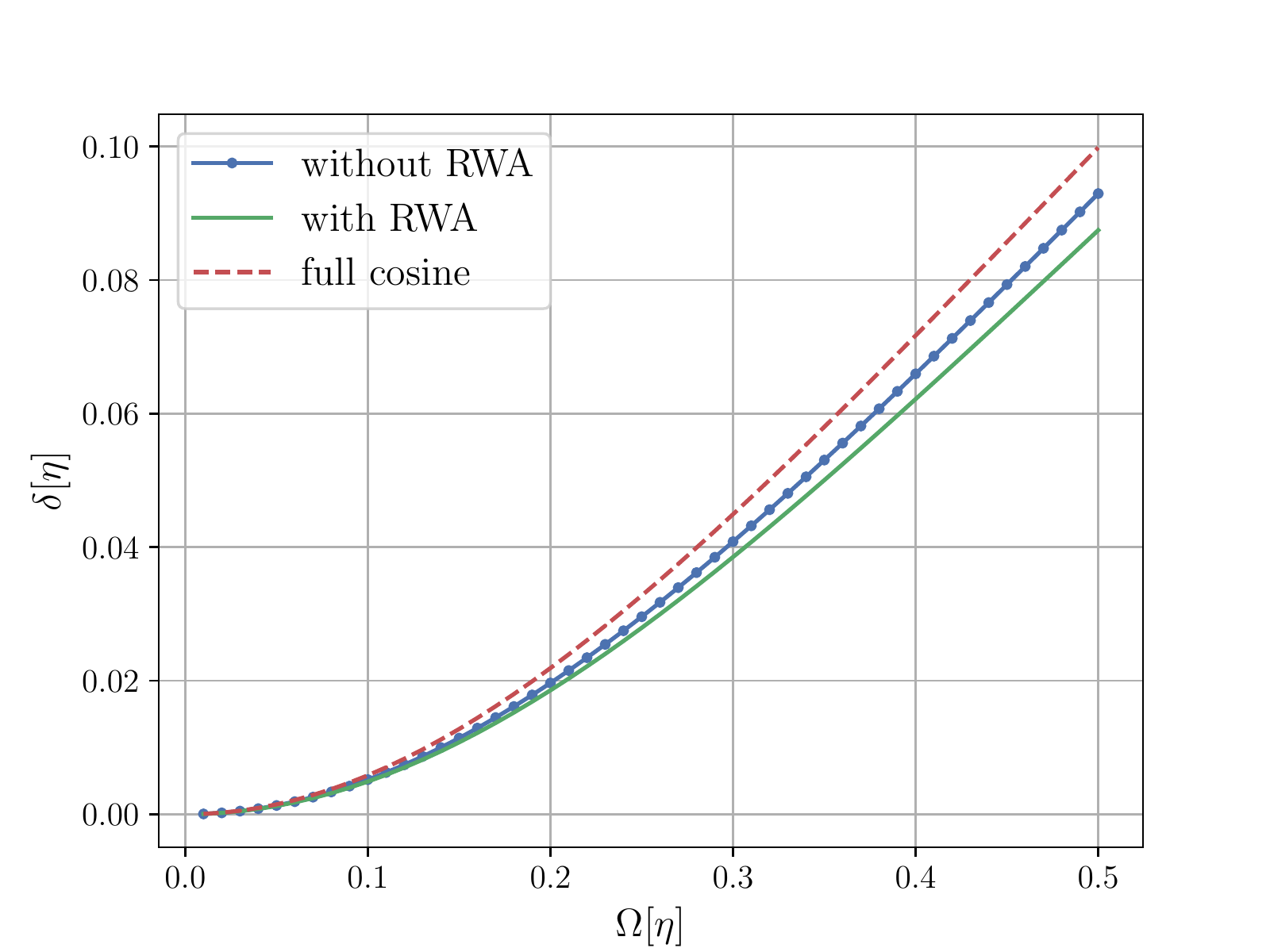}
\caption{$\delta(\Omega)$ that solves Eq.~(\ref{eq:insensativity}) for each of the three models listed in the legend. For the ``with RWA'' and ``without RWA'' models (where RWA stands for the rotating-wave approximation), we took 10 Fock
states. For the full cosine potential, we used  401 charge states, diagonalized the system, and then projected to the 10 lowest eigenstates. The error in the graph is less than $10^{-5}$. The system parameters are $\frac{\omega}{2\pi}=5$ GHz (corresponding to the transition frequency of the 2 first eigenstates of Eq.~(\ref{eq:full_cosine_Hamiltonian}) for the full cosine model)
and $\frac{\eta}{2\pi}=300$ MHz (corresponding to the transition frequency of the excited and doubly excited states of Eq.~(\ref{eq:full_cosine_Hamiltonian}) for the full cosine model).}
\label{fig: detuning_for_different_models}
\end{figure}

In this section, we describe the Hamiltonian that we used for our gate simulations.
Since we excite higher levels of the transmon, in order to examine
the properties of the spin-locked qubit correctly, we need to take into account
the full quantum description of the potential:
\begin{equation}
{H}_{fc}=4E_{C}{n}^{2}-E_{J}\cos({\phi})
\label{eq:full_cosine_Hamiltonian},
\end{equation}

where ${n}$ is the number of Cooper pairs that cross the Josephson
junction and ${\phi}$ is the magnetic flux in normalized units, which obey the canonical commutation relation $[\phi, n]=i$.
In Fig.~\ref{fig: detuning_for_different_models}, we compare $\delta(\Omega)$ that fulfills the clock condition in Eq.~(\ref{eq:insensativity}) for the Hamiltonian in Eq.~(\ref{eq:RWA_Hamiltonian}), for the same Hamiltonian without the rotating wave approximation,
\begin{equation}
H =\omega a^{\dagger}a-\frac{\eta}{2}a^{\dagger2}a^{2}-\Omega \sin({\omega_d t})\frac{(a-a^{\dagger})}{i},
\end{equation}
and
for the Hamiltonian in Eq.~(\ref{eq:full_cosine_Hamiltonian}) with a drive,
\begin{equation}
{H}_{fc}=4E_{C}{n}^{2}-E_{J}\cos({\phi}) -\Omega \sin({\omega_d t})n. 
\label{eq:driven_cosine_potential.}
\end{equation}

We note that, for the two models that are time-dependent, the time-independent dressed states $\{\ket{\tilde i}\}$ are not defined (without utilizing the rotating-wave approximation). Therefore, for these two models, we define the dressed eigenbasis as the (Floquet) eigenstates of $U(T)$, the unitary evolution for time $T=\frac{2\pi}{\omega_d}$ under the time-dependent Hamiltonian. 
This implies that all the numerical simulations (i.e., gates and ramp pulses) in these models project the state onto the eigenbasis only in integer multiples of the period $T=\frac{2\pi}{\omega_d}$. We see that, for sufficiently small Rabi frequencies, all
three models give the same result. However, for larger Rabi frequencies, the models deviate from each other as higher levels of the transmon get excited. 
The main difference between the full cosine model and the Kerr Hamiltonian [Eq.~(\ref{eq:RWA_Hamiltonian})] without the rotating wave approximation is that the matrix elements of the charge operator in the full cosine model do not exactly follow the harmonic oscillator pattern \cite{PhysRevA.76.042319}.  To account for the higher levels in the driven system, in all of the gate simulations and ramp pulses, the driven qubit was simulated with the Hamlitonian in Eq.~(\ref{eq:driven_cosine_potential.}).

We should note that, for the chosen parameters, the deviations in $\delta(\Omega)$ between the models are on the order of a few MHz or less.  In an experimental setting, similar deviations in the detuning are expected due to spurious couplings and parameter drifts. While we continue---in the one-qubit-gate, two-qubit-gate, and ramp simulations---using the full cosine potential and the Floquet basis, in practice, the detuning should be scanned to maximize the dephasing time, and the clock condition from Eq.~(\ref{eq:RWA_Hamiltonian}) likely provides a sufficiently good starting point for such a scanning procedure.

\section{one-qubit gate}
\label{section: one-qubit gate}

\begin{figure}[t]
\centering
\includegraphics[width=8.6cm]{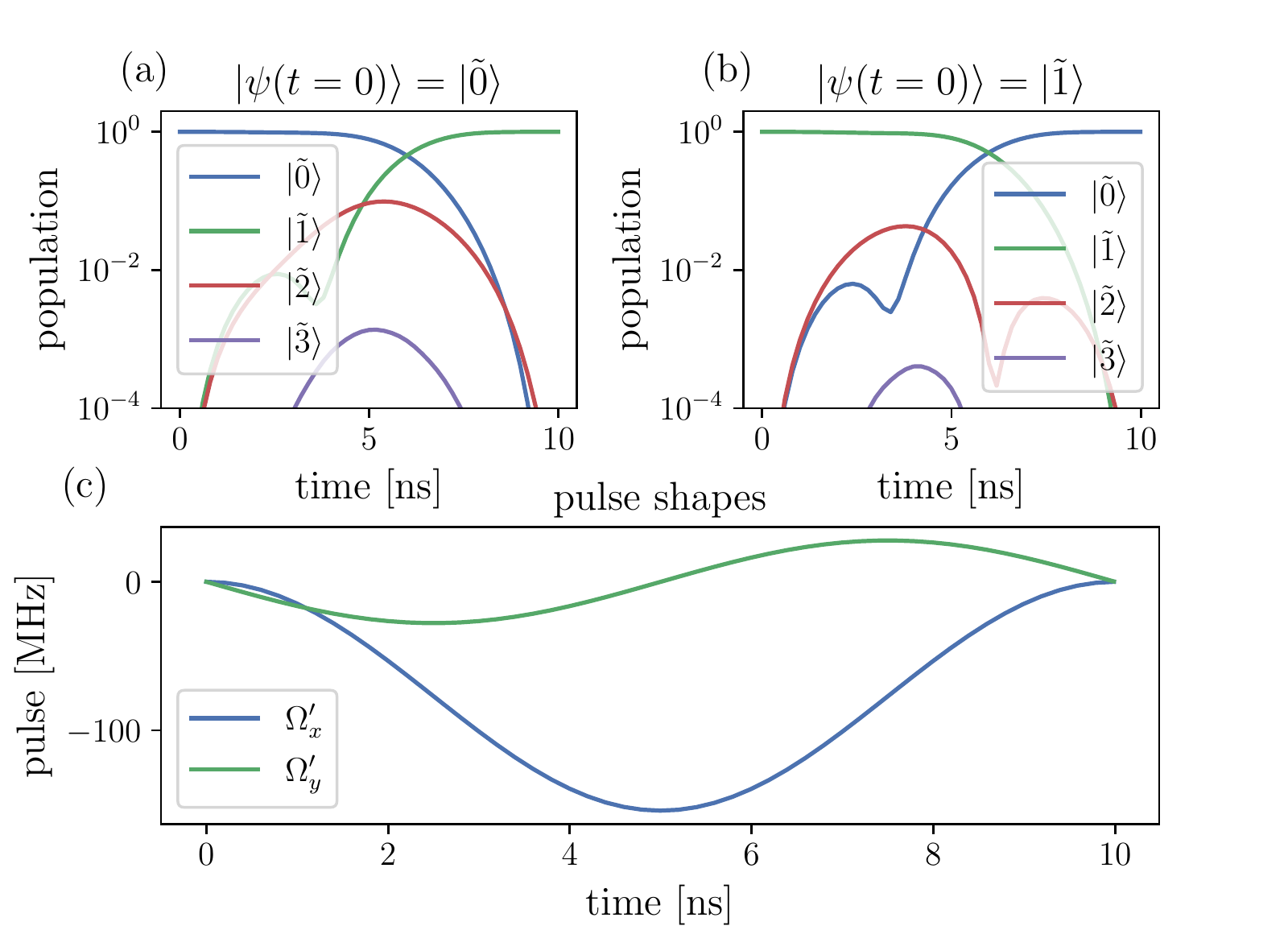}
\caption{$X$ gate for the dressed qubit with $\Omega = 0.2 \eta$. (a) and (b) show populations of the dressed (Floquet) eigenstates. Although the points were taken only at integer multiples of $\frac{2\pi}{\omega_d}$, the graphs show smooth curves for readability. The initial state is (a)  $|\tilde{0}\rangle$ or (b) $|\tilde{1}\rangle$. (c) shows the pulse shape, which is explained in Eq.~(\ref{eq:1_q_gate_pulse_shape}).
The parameters for this simulation are $\frac{\omega}{2\pi}=5$ GHz, $\frac{\eta}{2\pi}=300$ MHz, $A=-1.7\pi$, $l_{D}=-1.084$, and $\frac{\omega_{d}'}{2\pi}=5031.485$ MHz.
$\delta$ is chosen to be on the $\delta(\Omega)$ curve. The gate time is $T_g=10$ ns, and the average infidelity is $3\cdot10^{-6}$.}
\label{fig:one-qubit-gate-main-text}
\end{figure}

To implement a one-qubit rotation, we use the DRAG scheme \cite{PhysRevLett.103.110501, PhysRevA.83.012308}. One of the challenges in designing one-qubit gate is that there are no selection rules when using microwave pulses that couple to the charge operator: even in the limit $\Omega \rightarrow 0$, the annihilation operator in the dressed basis is highly non-trivial: $a = \frac{1}{2}[ (|\tilde 0\rangle\langle \tilde 0| - |\tilde 1\rangle\langle \tilde 1|)+(|\tilde 1\rangle\langle \tilde 0|+|\tilde 0\rangle\langle \tilde 2|-h.c) ]$.
See Appendix \ref{appendix:Eigenenergies and eigenstates of the SL qubit} for order-$\epsilon$ and order-$\epsilon^2$ corrections to this expression for $a$. Nevertheless, since DRAG is designed to avoid leakage, we found this method quite robust, even for the spin-locked qubit.
We start from the Hamiltonian with the full cosine potential with a drive [Eq.~(\ref{eq:driven_cosine_potential.})].
To induce rotations on the dressed qubit's Bloch sphere, we add to the Hamiltonian in Eq.~(\ref{eq:driven_cosine_potential.}) the two DRAG terms
\begin{equation}
\left(-\Omega'_{x}(t)\sin(\omega'_{d}t)+\Omega'_{y}(t)\cos(\omega'_{d}t)\right)n,
\end{equation}
where we used the pulse shape
\begin{align}
\Omega_{x}'(t) & =\frac{A}{T_{g}\tilde{n}_{01}}(1-\cos(2\pi t/T_{g})),\nonumber\\
 & \Omega_{y}'(t)=-\frac{l_{D}}{\eta}\cdot\frac{d}{dt}\Omega_{x}'(t),
\label{eq:1_q_gate_pulse_shape}
\end{align}
where $T_{g}$ is the gate time. $\tilde{n}_{01}=\bra{\tilde{0}}n\ket{\tilde{1}}$ is the matrix element of the charge operator between the spin-locked qubit states (this matrix element was introduced into the definition of $\Omega'_x$ to follow the convention of the original DRAG formulation \cite{PhysRevLett.103.110501, PhysRevA.83.012308}). $A$, $l_{D}$, and the drive frequency $\omega_{d}'$ are free parameters for the optimization. For $\Omega=0.2\eta$, we obtain an $X$ gate with average infidelity of $3 \cdot10^{-6}$ 
\cite{NIELSEN2002249_revised}, for a gate time $T_{g}=10$ ns. Figure \ref{fig:one-qubit-gate-main-text} shows population graphs and the pulse shape of the gate. In order to show that our gate can be realized with higher Rabi frequencies, we also simulated the gate for $\Omega=0.3\eta$. Additionally, we simulated the $X_{\frac{\pi}{4}}=e^{i\sigma_x \frac{\pi}{4}}$ gate, where $\sigma_x$ is the Pauli X matrix, for $\Omega=0.2\eta$. These two results have similar gate times and infidelities to the gate presented above and can be seen in Appendix \ref{appendix: one-qubit gate}. We note that, since the gate is fast, we expect that noise on the spin locking drive amplitude or the additional drive will not limit the gate fidelity under standard assumptions on the noise [see Sec.~\ref{subsec: limiting factors}].

\setlength\lineskip{0pt}

\section{two-qubit gate}
\label{section: two-qubit gate}

\begin{figure}[t]
\centering
\includegraphics[width=8.6cm]{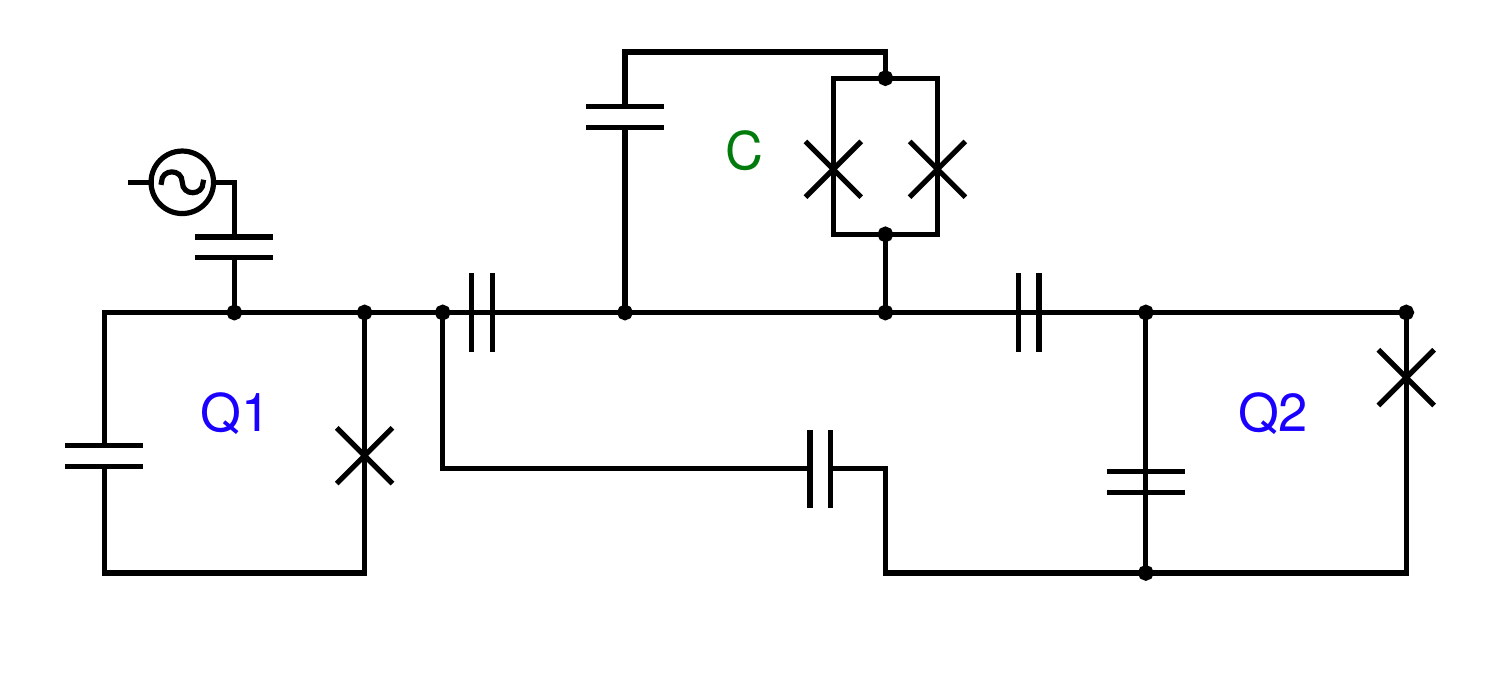}
\caption{Bus-below-qubit architecture. Q1 and Q2 are transmons (could be fixed or tunable) that serve as the qubits, while C is a tunable coupler. Q1 is driven continuously.}
\label{fig:BBQ}
\end{figure}

The main difficulty in using spin locking with superconducting qubits is the realization of a two-qubit gate. There are two challenges. First, the spin-locked qubit should stay inside the computational basis, since only the first two levels are protected. Second, there are no selection rules, as discussed in Sec.~\ref{section: one-qubit gate}. This means that, if we couple two transmons transversely, the charge-charge interaction would result in many coupled dressed levels, making it difficult to target a specific transition. 
An adiabatic $ZZ$ gate in the bus-below-qubit architecture \cite{PhysRevLett.127.080505} [see Fig.~\ref{fig:BBQ}] overcomes both of these challenges by staying within the computational basis and not relying on targeting a specific transition. Another advantage of the adiabatic ZZ gate is its relative robustness to multi-body error terms arising from cross-talk, as we show in Appendix \ref{appendix:ZZ cross-talk}. We choose to spin-lock one qubit only, for reasons that will be explained below. In the adiabatic ZZ gate, by controlling the flux on the coupler, we create an effective two-qubit Hamiltonian, in a frame that rotates with the qubit frequencies:
\begin{equation}
    H_{eff}=\tilde{g}_{zz}(t)\tilde{\sigma}_z^1\sigma_z^2,
\end{equation}

where $\tilde{\sigma}_z^1$ is the Pauli $Z$ matrix in the dressed basis of the driven qubit 1 ($q_1$), while $\sigma_z^2$ is the Pauli $Z$ matrix of the un-driven qubit 2 ($q_2$).  We define $\tilde{g}_{zz}(t)$ as
\begin{equation}
    \tilde{g}_{zz}(t)=\frac{1}{4}\{E_{\tilde{0}0}(t)-E_{\tilde{0}1}(t)-E_{\tilde{1}0}(t)+E_{\tilde{1}1}(t)\},
\end{equation}

where $E_{\tilde{i}j}(t)$ is the instantaneous (Floquet) energy for the dressed qubit in state $|\tilde{i}\rangle$, the second qubit in state $|j\rangle$, and the coupler is in state $\ket{0}$.  
To implement the unitary $U=e^{-i\frac{\pi}{4}\sigma_z^1\sigma_z^2}$, which is equivalent to a $CZ$ gate up to single-qubit $Z$ rotations, we require $\int_0^{T_g}\tilde{g}_{zz}(t)dt=\frac{\pi}{4}$, where $T_g$ is the time of the gate.

\begin{figure}[t]
\centering
\includegraphics[width=8.6cm]{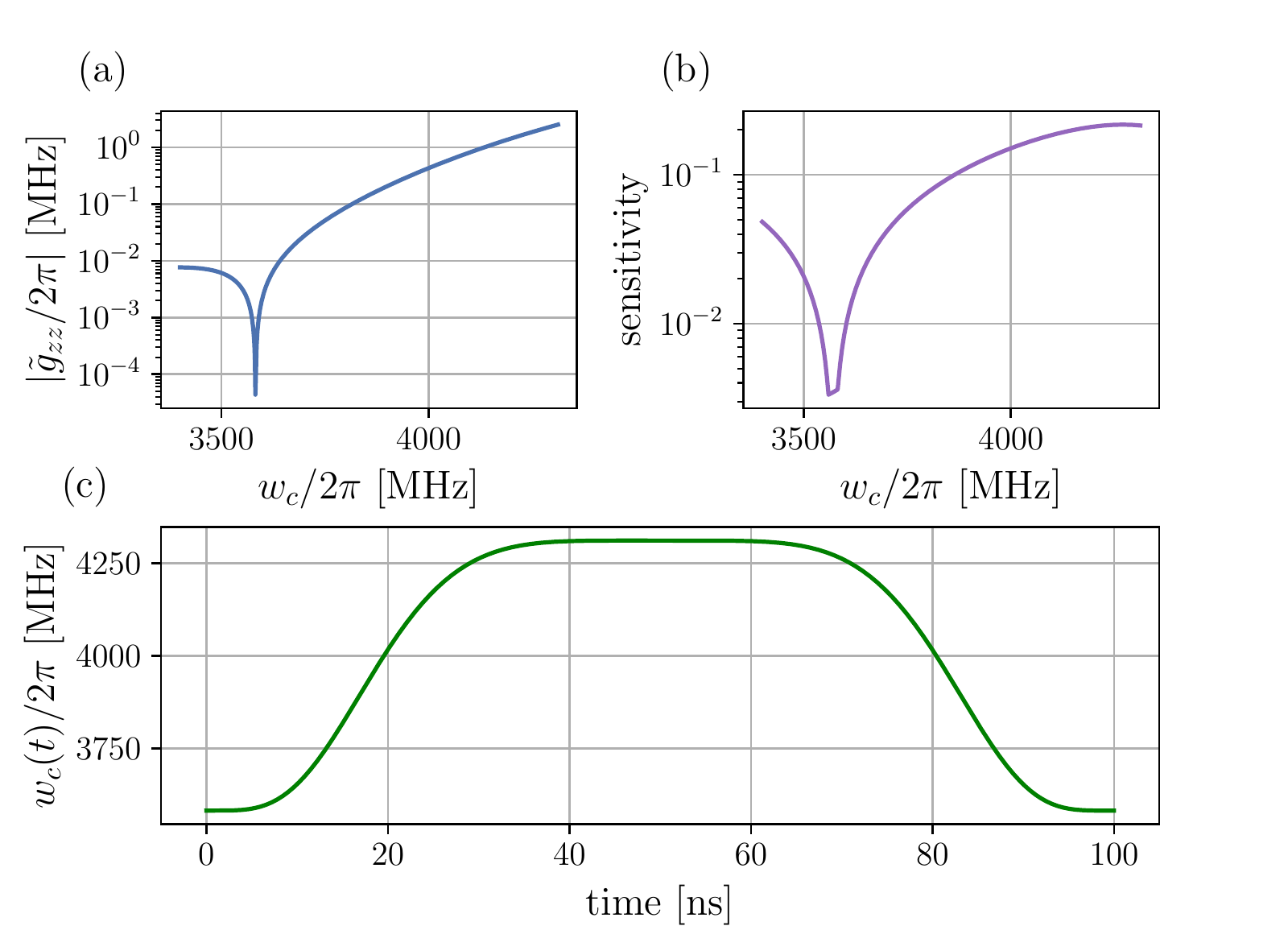}
\caption{(a) $\frac{\tilde{g}_{zz}}{2\pi}$ as function of $\frac{\omega_c}{2\pi}$. (b) The phase-noise sensitivity as a function of $\frac{\omega_c}{2\pi}$. For a thorough discussion of the sensitivity measure in a hybridized system, see Appendix \ref{appendix: two-qubit gate - sensativity}. (c) The frequency of the coupler as a function of time during the two-qubit gate.}
\label{fig:g_zz for coupler and pulse shape}
\end{figure}

The Hamiltonian of the bus-below-qubit with spin-locking is
\begin{align}
H & ={H}_{fc,1}-\Omega\cdot \sin(\omega_{d}t){n}_{1}+g_{1c}{n}_{1}\frac{c-c^{\dagger}}{i}\nonumber \\
 & + \omega_{2}q_{2}^{\dagger}q_{2}-\frac{\eta_{2}}{2}q_{2}^{\dagger2}q_{2}^{2}+g_{2c}\frac{(q_{2}-q_{2}^{\dagger})}{i}\frac{(c-c^{\dagger})}{i}\nonumber \\
 & +\omega_{c}(t)c^{\dagger}c-\frac{\eta_{c}}{2}c^{\dagger2}c^{2}+g_{12}{n}_{1}\frac{(q_{2}-q_{2}^{\dagger})}{i},
 \label{eq:2q_Hamiltonian}
\end{align}
where ${H}_{fc,1}$ is the Hamiltonian with the full cosine potential for qubit 1 (the driven qubit) and ${n}_{1}$ is the charge operator. $q_{2}$ ($c$) is the second-qubit (coupler) annihilation operator. The two systems that are not driven are described as anharmonic oscillators since we do not excite their higher levels. We require our system to have a ZZ-free point ($\tilde{g}_{zz}=0$) where we could park the coupler, and adiabatically tune the coupler frequency $\omega_c(t)$ to increase $|\tilde{g}_{zz}|$ to achieve a fast CZ gate, without encountering any avoided crossings. Although spin locking reduces the direct, second-order, ZZ interaction between the two qubits, there still could be a relatively strong ZZ that originates from the coupler and manifests as a third- or fourth-order effect.

\subsection{Pulse shape and gate parameters}
\label{subsec: 2q - Pulse shape and gate parameters}
In our simulations, we choose the spin-locked qubit parameters  $\frac{\omega}{2\pi} =5$ GHz [corresponding to the transition frequency between the lowest two eigenstates of Eq.~(\ref{eq:full_cosine_Hamiltonian})] and $\frac{\eta}{2\pi} =300$ MHz [corresponding to $2 \omega$ minus the transition frequency between the ground state and the second excited state of Eq.~(\ref{eq:full_cosine_Hamiltonian})], the drive Rabi frequency $\Omega=0.2\eta$, the second-qubit parameters $\frac{\omega_{2}}{2\pi} =4.55$ GHz and $\frac{\eta_{2}}{2\pi} =200$ MHz, and the coupler anharmonicity  $\frac{\eta_{c}}{2\pi}=200$ MHz. The couplings are $\frac{g_{1c}}{2\pi}=190$ MHz, $\frac{g_{2c}}{2\pi}=200$ MHz, and $\frac{g_{12}}{2\pi}=-30$ MHz. We use the following pulse shape to control the frequency of the coupler:
\begin{align}
&
\omega_{c}(t)= \nonumber \\
&
\begin{cases}
\omega_{s}+\omega_{fs}\left(1-\left(1-\sin(\frac{\pi t}{2T_{r}})\right)^{m}\right)^{n}& t<T_{r}\\
\omega_{f} & T_{r}\leq t\leq T_{fr}\\
\omega_{s}+\omega_{fs}\left(1-\left(1-\sin(\frac{\pi(t-T_{f})}{2T_{r}})\right)^{m}\right)^{n} & t>T_{fr}
\end{cases},
\label{eq:2q_pulse_shape}
\end{align}

where $\omega_s$ is the coupler frequency during idling (i.e, the $ZZ$-free point), $T_f$ is the flat-top time, and $T_g = T_f + 2 T_r$ is the total gate time. The ramp time $T_r$, the final frequency $\omega_f$, and the shaping parameters $m$ and $n$ are free parameters for the optimization. $T_{fr}=T_f+T_r$ and $\omega_{fs}=\omega_f-\omega_s$. 
We set the gate time to be $T_g = 100$ ns. The parameters that were found by the optimization are $\frac{\omega_f}{2\pi}=4310.552$ MHz, $T_r=49.975$ ns (corresponding to $T_f = 0.05$ ns, which makes it negligible in this realization, and we keep it for generality), $m=2.554$, and $n=4.716$.  Figure \ref{fig:g_zz for coupler and pulse shape}(a) shows $|\frac{\tilde{g}_{zz}}{2\pi}|$ for different coupler frequencies, while Fig.~\ref{fig:g_zz for coupler and pulse shape}(b) shows the phase-noise sensitivity of qubit 1. Thus, if we set $\frac{\omega_{c}}{2\pi} \approx 3.58$ GHz, we effectively get $\tilde{g}_{zz}\approx0$ and minimize the sensitivity, thus $\frac{\omega_{s}}{2\pi} \approx 3.58$. By tuning the coupler to $\frac{\omega_{c}}{2\pi}\approx 4.31$ GHz [Fig.~\ref{fig:g_zz for coupler and pulse shape}(c)], we achieve $|\frac{\tilde{g}_{zz}}{2\pi}|\approx 2.5$ MHz, without crossing any resonances along the adiabatic path, which would have manifested themselves as dips in Fig.~\ref{fig:g_zz for coupler and pulse shape}(b). Moreover, spin locking enables us to work outside of the straddling regime \cite{Krantz_2019_revised} and still effectively cancel the ZZ interaction. Our CZ gate has an average infidelity of
$2\cdot10^{-7}$ (which in practice means that the gate is coherence limited). Most of the infidelity comes from leakage, which is in turn related to the adiabaticity of the process set by the gate time. The population graph for the gate can be found in Appendix \ref{appendix: two-qubit gate - population}. We note that our gate fidelity is mainly $T_1$-limited. For transmons with $T_1=100 \mu s$, we should expect gate fidelity of $\approx \frac{T_g}{T_1}=10^{-3}$. Since we expect that noise on the spin locking drive amplitude will contribute to the fidelity quadratically [see Sec.~\ref{subsec: limiting factors}], it will not limit the gate fidelity. Furthermore, in Sec.~\ref{section: Phase noise and data vs.~ancilla errors}, we show that the gate is protected from dephasing, including dephasing of the coupler, which means that noise on the flux drive will not limit the gate fidelity. 
 
\subsection{Gates with different parameters}
\label{subsec: 2q - additional gates}

We note that small variations in the couplings and transmon frequencies due to fabrication errors would yield different ZZ curves [Fig.~\ref{fig:g_zz for coupler and pulse shape}(a)] and thus mainly impact the gate speed. In addition, in our scheme, we demand $\tilde{g}_{zz}=0$ at the resting (idling) point of the coupler. This requirement can be relaxed if we allow ZZ interaction during idling that is sufficiently small to have a negligible contribution to the overall infidelity compared to $T_1$. This compromise can help in choosing gate parameters. As an example, we simulated the gate with $\frac{g_{1c}}{2\pi}=\frac{g_{2c}}{2\pi}=150$ MHz (smaller, and hence potentially easier to realize experimentally, than $\frac{g_{1c}}{2\pi}=190$ MHz and $\frac{g_{2c}}{2\pi}=200$ MHz used in the previous section) and $\frac{g_{12}}{2\pi}=-5$ MHz. The rest of the parameters are the same as described in Sec.~\ref{subsec: 2q - Pulse shape and gate parameters}. Using these parameters, we realized a CZ gate (up to single-qubit Z rotations) with average infidelity $\approx 2\cdot 10^{-5}$ (although this infidelity is two orders of magnitude larger than for the parameters for the gate in Sec.~\ref{subsec: 2q - Pulse shape and gate parameters}, once $T_1$ is taken into account, it will dominate the infidelity and both gates will perform similarly). As in the gate of Sec.~\ref{subsec: 2q - Pulse shape and gate parameters}, most of the infidelity is attributed to non-adiabaticity errors. We used the same pulse shape as in Eq.~(\ref{eq:2q_pulse_shape}) with $T_g=100$ ns.

Additionally, one may ask how fast can we perform this gate. As we argued above, the main fidelity limitation of the gate is due to $T_1$ processes. Therefore, we can reduce the time of the gate down to the point where the infidelity due to non-adiabatic errors is comparable to the infidelity from $T_1$. As an example, we simulated the gate using the same couplings as in Sec.~\ref{subsec: 2q - Pulse shape and gate parameters} for $T_g=80$ ns, with average infidelity of $3\cdot 10^{-4}$, which is comparable to infidelity from $T_1$ during the gate time with current state-of-the-art transmons.
 Details regarding the optimized parameters for the two additional gates are given in Appendix \ref{appendix: two-qubit gate - realizations}, and the population graphs are in Appendix \ref{appendix: two-qubit gate - population}.

\section{Phase noise and data vs.~ancilla errors}
\label{section: Phase noise and data vs.~ancilla errors}

We have just discussed a CZ gate with only one of the qubits spin-locked, thus leaving the second qubit unprotected from dephasing. We propose using our CZ gate so that the spin-locked qubit ($q_1$) is a data qubit in a surface code architecture, while the second qubit ($q_2$) is an ancilla. This offers two advantages:  lower experimental complexity in (a) driving only one qubit and (b) making frequent ancilla measurements in the computational (not dressed) basis. Since ancilla phase noise, which couples to $Z_2$, commutes with the CZ gate, the resulting $Z_2$ errors will not propagate from the ancilla to the data before and after the gate. Thus, such $Z_2$ errors can only result in measurement errors. However, $Z_2$ errors could potentially propagate to the data qubit during the gate itself. To understand this propagation, we simulate the gate while adding a constant shift $\zeta q_{i}^\dagger q_{i}\:,i\in \{1,2\}$ to the Hamiltonian. For each $T_s= \frac{2\pi}{\zeta}$, we calculate the average infidelity and extract the underlying Pauli channel under the Pauli-twirl approximation 
\cite{PhysRevA.54.3824, doi:10.1126/science.1145699_revised, PhysRevA.78.012347, PhysRevA.72.052326, PhysRevA.86.062318, PhysRevA.88.012314, Katabarwa2015_revised}, see Fig.~\ref{fig:t_phi_and_threshold}(a) and (c).

\subsection{Protection of the one-qubit gate}

For the one-qubit $X$ gate (The same parameters as in Fig.~\ref{fig:one-qubit-gate-main-text}), Fig.~\ref{fig:t_phi_and_threshold}(a) shows that the shift causes only a small error (even for $T_s=0.2\: \mu s$). We also plot the line $P_z(T_s)=\left(\pi\frac{T_g}{T_s}\right)^2$, which is the expected Z error probability under the Pauli-twirl approximation without spin locking, see Appendix \ref{appendix: PTA_for_z_errors}. We see that $Pr(Z)$ is $\approx 14$ times smaller than $P_z(T_s)$. Thus, the one-qubit gate is robust against such noise. We note that, during the gates, state $\ket{\tilde{2}}$, which is not protected by the clock condition, is also populated. Nevertheless, since the gate is fast, and the population outside the computational basis is $\approx 10\%$, dephasing of the transmon will not affect the fidelity of the gates significantly, as shown above.

\subsection{Protection of the two-qubit gate}

\begin{figure}[t]
\centering
\includegraphics[width=8.6cm]{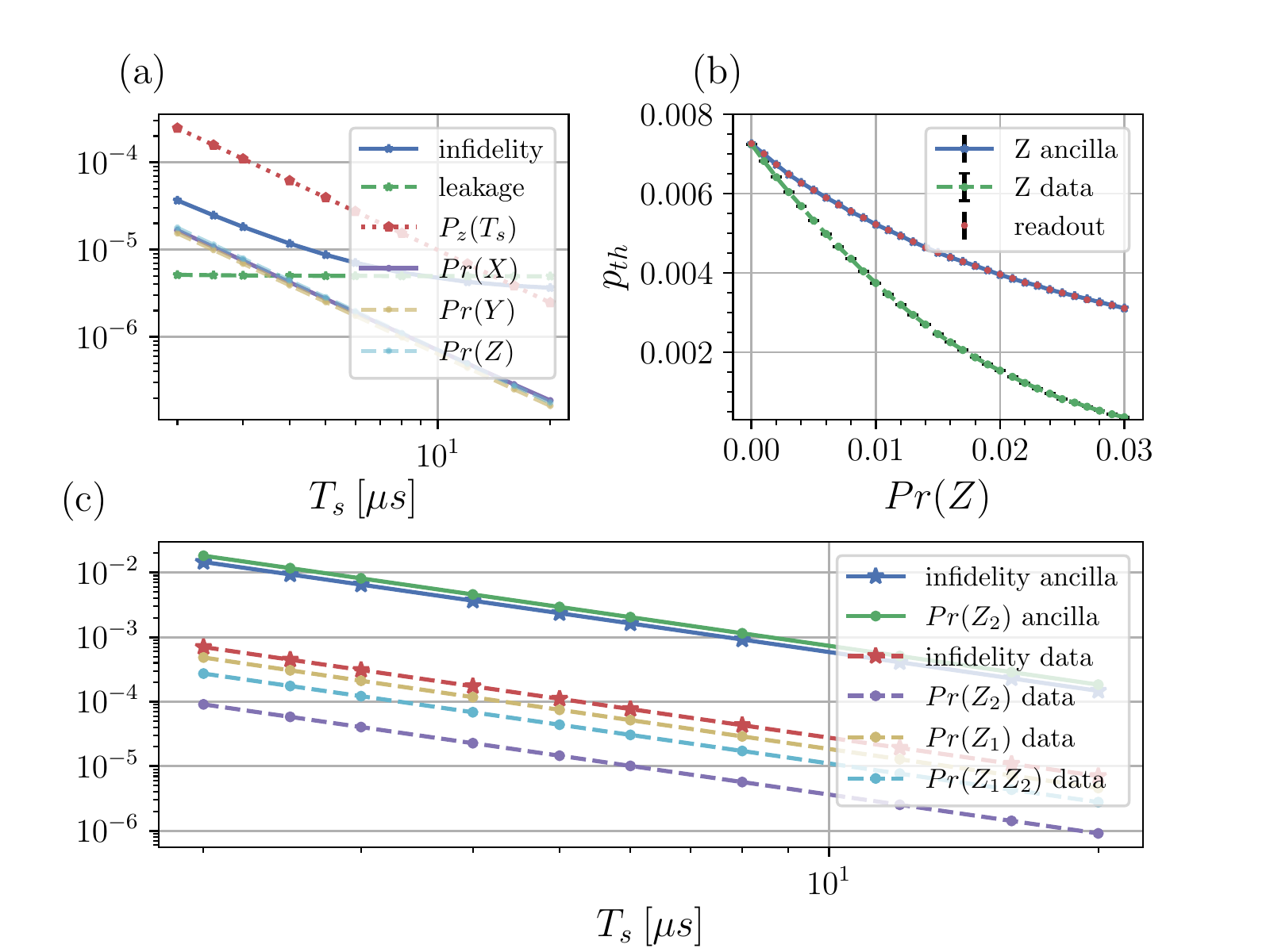}
\caption{The average infidelity and the Pauli probabilities as a function of $T_{s}$ for a one-qubit $X$ gate (a) and a two-qubit CZ gate (c). We define leakage as $1-\sum_{p\in{\mathcal{P}}}{Pr(p)}$, for $\mathcal{P}$ the Pauli group. 
$\sum_{p\in{\mathcal{P}\setminus I}}{Pr(p)}+\textrm{leakage}$ is a rough estimate for the average infidelity, since the latter also accounts for unitary errors.
(a) one-qubit $X$ gate with $\Omega=0.2\eta$ and $T_g=10$ ns. $P_z(T_s)$ is the Z error probability without spin locking.
(c) two-qubit CZ gate. The gate parameters appear in Sec.~\ref{subsec: 2q - Pulse shape and gate parameters}. The curves where the shift was added to the data (ancilla) qubit are named "data" ("ancilla"). (b) The threshold for the depolarizing channel in surface code simulations for different probabilities of added
Z errors for three models: Z errors on the ancilla (blue) and on the data (green), and the equivalent measurement error (red). For more details on (b), see Appendix \ref{appendix: threshold}.}
\label{fig:t_phi_and_threshold}
\end{figure}

For the two-qubit CZ gate (The same parameters as in Sec.~\ref{subsec: 2q - Pulse shape and gate parameters}), Fig.~\ref{fig:t_phi_and_threshold}(c) shows that, when we shift the ancilla, almost all of the infidelity can be explained by $Z_2$ errors. When we shift the data, we see that the infidelity is $\approx 25$ times lower than in the ancilla case, meaning that spin locking significantly improves data qubit robustness to phase noise. In Appendix \ref{appendix:coupler_dephase}, we show that, because the hybridization of the coupler with the other qubits is small, adding dephasing to the coupler has an insignificant effect on the gate. 

\subsection{Phase noise in the surface code}
Lastly, we show that the surface code is more robust to ancilla $Z$ errors than to data $Z$ errors. We simulate the surface code using Stim \cite{gidney2021stim} and extract the threshold for three different models. In all of the models, we apply a depolarizing channel after every one-qubit operation and before every measurement, and a two-qubit depolarizing channel after every CZ, with the same error rate. For the ``$Z$-ancilla'' (``$Z$-data'') model, we also apply after each CZ a Pauli $Z$ with probability $Pr(Z)$ to the ancilla (data) qubit. In the ``readout'' model, before each ancilla measurement, we apply a Pauli $X$ with probability $p_{readout}=4Pr(Z)(1-Pr(Z))^3+4Pr(Z)^3(1-Pr(Z))$, to account for both cases of one or three $Z$ errors that lead to a measurement error. Figure \ref{fig:t_phi_and_threshold}(b) shows that the ``$Z$-ancilla'' model and the ``readout'' model give the same result, while the ``$Z$-data'' model has a much lower threshold (an order of magnitude lower than the other models for $Pr(Z)\approx 0.03$). Thus, even if the ancilla qubit experiences strong phase noise, the code would still have a reasonable threshold with respect to other sources of noise. 

\section{State preparation and measurement in the dressed basis}
\label{section: State preparation and measurement in the dressed basis}
We propose preparing (measuring) the dressed states by using standard preparation (measurement) in the computational basis and ramping up (down) the spin-locking drive to dress (undress) the qubit state. Another option, which we discuss in Appendix \ref{appendix:measure}, is to measure the dressed qubit while it is still driven \cite{PhysRevLett.117.133601, PhysRevLett.120.040505}. To engineer efficient ramps between the undressed and dressed computational basis states, we use Krotov's method \cite{GoerzSPP2019}. We used an ansatz of a DRAG-inspired pulse as a starting point for Krotov's optimization. The control Hamiltonian is taken to be of the form
\begin{equation}
\left(-\Omega_{x}(t)\sin(\omega_{d}t)+\Omega_{y}(t)\cos(\omega_{d}t)\right)n.
\label{eq:krotov}
\end{equation}

We define $T_r$ as the ramp time. During ramp-up, $\Omega_{x}(0)=\Omega_{y}(0) = 0$, $\Omega_{x}(T_r)=\Omega$, and $\Omega_{y}(T_r)=0$. During ramp-down, the initial and final conditions are reversed. Here $\Omega$ and $\omega_d$ are the final values for our spin-locking continuous drive. Using these pulses, we reached average infidelity below $10^{-4}$ with a $50$ ns ramp time. The population graphs during the pulse can be seen in Appendix \ref{appendix:ramp}. 

Since optimal coherent control methods could generate pulse envelopes with high frequencies, it is possible to increase the time of the ramp pulse and achieve smoother pulses. For example, with $T_r=200$ ns, we can map between the bare and the dressed computational basis with infidelity below $10^{-4}$, see Appendix \ref{appendix:ramp} for the population graphs. That, in turn, will make the qubit more sensitive to phase noise during the ramp. However, since we suggest driving only the data qubits in a surface code architecture, these errors are important only for state preparation and measurement, at the beginning and the end of the full quantum error correction experiment (or algorithm). For the state preparation, any such error could be detected using an error detection cycle of the surface code (before starting the algorithm). For the measurement, since we suggest doing a ramp down and then a standard measurement along the bare $\sigma_z$ axis, phase noise will not affect the final result. $T_1$ processes during the ramp down would increase the measurement error for longer ramp time, but we expect it to be comparable to gate errors using current state-of-the-art hardware.

\section{Leakage detection and conversion to erasure}
\label{section: Leakage detection with the SL qubit}

\begin{figure}[t]
\centering
\includegraphics[width=8.6cm]{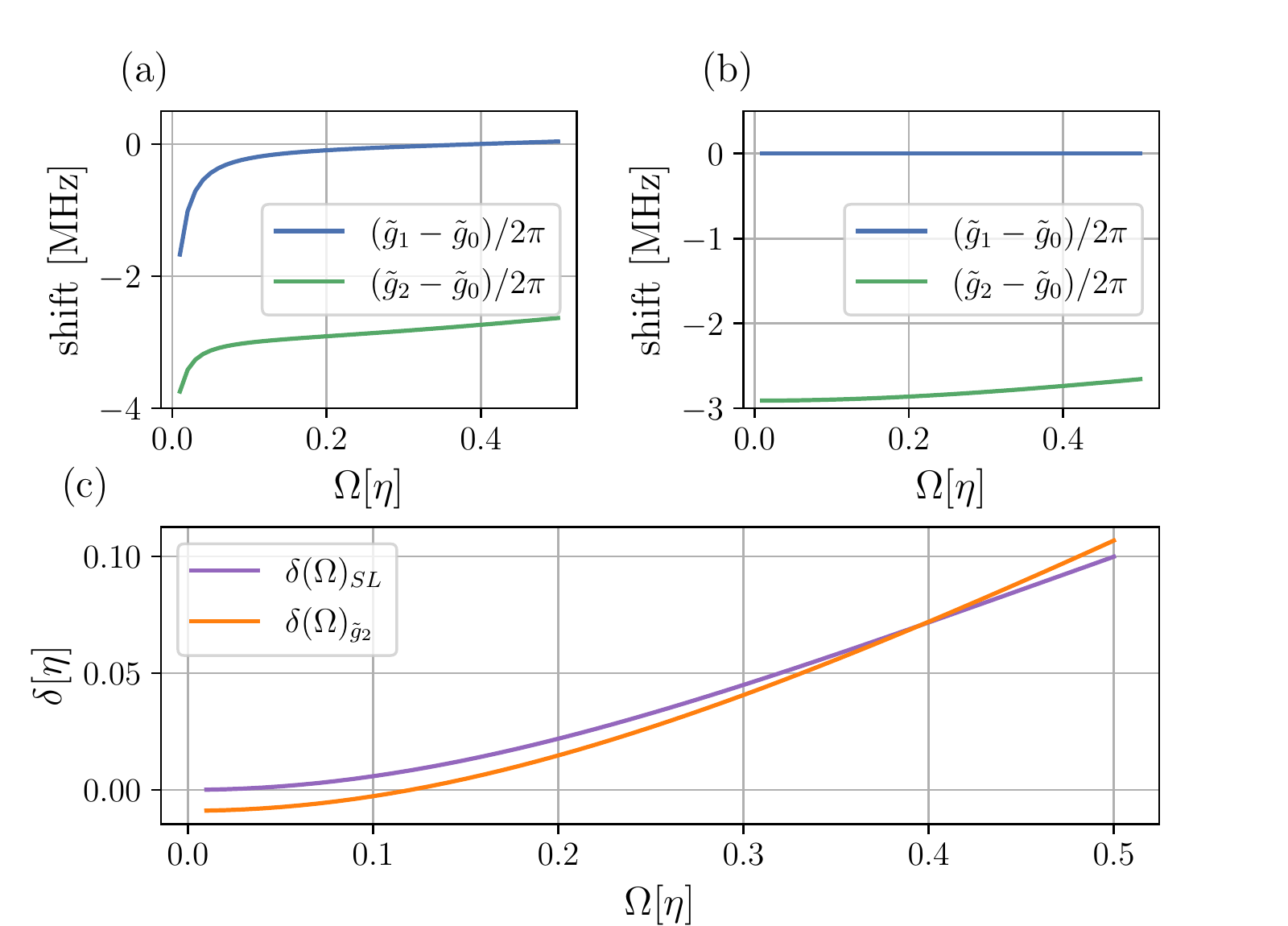}

\caption{Differences between resonator coupling strengths for (a) the standard $\delta(\Omega)$ satisfying the clock condition [Eq.~(\ref{eq:insensativity})]  and (b) when optimizing $\delta$ for minimal $|\tilde{g}_1-\tilde{g}_0|$. (c) $\delta(\Omega)_{SL}$ and $\delta(\Omega)_{\tilde{g}_2}$ as a function of $\Omega$, without the Stark shift from the resonator $\frac{g^2}{w-\omega_b}$.}
\label{fig:leakgae}
\end{figure}

In Sec.~\ref{section: Spin locking of a transmon}, we described how the spin-locked qubit is protected from the fluctuations of the population in a dispersively coupled resonator. We can further exploit this property by engineering a small dispersive shift of the resonator when the spin-locked qubit is in the computational basis and a strong shift when the spin-locked qubit leaked to the $|\tilde{2}\rangle$ state. 
In the dispersive regime, we have the Hamiltonian
\begin{equation}
    H=\sum_i \tilde{E}_i |\tilde{i}\rangle \langle \tilde{i}| + \omega_b b^\dagger b + \sum_i\tilde{g}_i|\tilde{i}\rangle \langle \tilde{i}| b^\dagger b,
\end{equation}

where $b$ is the annihilation operator for the resonator, $\tilde{E}_i$ are the eigenenergies of the spin-locked transmon, and $|\tilde{i}\rangle$ are the eigenstates of the spin-locked transmon. $\tilde{g}_i$ are the cross-Kerr shifts corresponding to different levels of the spin-locked transmon. 
We simulated this system with the following parameters: the qubit frequency is $\frac{\omega}{2\pi}=5$ GHz and the resonator frequency is $\frac{\omega_b}{2\pi}=6.1$ GHz. The qubit anharmonicity is  $\frac{\eta}{2\pi}=300$ MHz. The drive detuning is set to be $\delta=\delta(\Omega)+\frac{g^2}{w-\omega_b}$, to account for the Stark shift from the resonator. To make sure that the contribution from higher levels of the spin-locked qubit is taken correctly into account, the simulation was done using the full cosine potential. Figure \ref{fig:leakgae}(a) shows $\tilde{g}_i-\tilde{g}_0$ for $i\in\{1,2\}$. We see that, although spin locking reduces significantly the relative shift between levels $\ket{\tilde{0}}$ and $\ket{\tilde{1}}$ (i.e., $\tilde{g}_1-\tilde{g}_0\approx 0$), we still have a relatively strong shift for level  $\ket{\tilde{2}}$. In other words, $|\tilde{g}_2-\tilde{g}_i|\gg |\tilde{g}_1-\tilde{g}_0|$ for $i\in\{0,1\}$. Thus, the resonator could potentially distinguish between the qubit subspace and the leakage subspace, but not within the qubit subspace. This feature could be used to detect leakage, by sending a pulse into the resonator that is resonant with $\omega_b+\tilde{g}_2$.

Another option, which we show in Fig. \ref{fig:leakgae}(b), is to choose $\delta$ to completely nullify $\tilde{g}_1-\tilde{g}_0$, for any given $\Omega$ \cite{PhysRevX.13.011008}. This way, we could have potentially perfect leakage detection for any Rabi frequency. For this section only, we denote this $\delta$ as $\delta(\Omega)_{\tilde{g}_2}$ and the standard $\delta$ that reduces sensitivity to phase noise as $\delta(\Omega)_{SL}$. Figure \ref{fig:leakgae}(b) shows that we can indeed find a $\delta$ (i.e.~$\delta(\Omega)_{\tilde{g}_2}$) that nullifies $\tilde{g}_1-\tilde{g}_0$ while still resulting in a strong shift when the spin-locked qubit is in state $\ket{\tilde{2}}$. Figure~\ref{fig:leakgae}(c) shows $\delta(\Omega)_{SL}$ and $\delta(\Omega)_{\tilde{g}_2}$. We should note that using $\delta(\Omega)_{\tilde{g}_2}$ means that the spin-locked qubit is less protected from phase noise, and therefore there is typically a trade-off between phase-noise protection and the ability to detect leakage. 
An interesting feature that can be seen both in Fig.~\ref{fig:leakgae}(b) and in Fig.~\ref{fig:leakgae}(c) is that, for $\Omega\approx0.4\eta$, $\delta(\Omega)_{\tilde{g}_2}=\delta(\Omega)_{SL}$ and there is thus no trade-off.

Therefore, we can potentially use the leakage detection as a heralded erasure error in the surface code \cite{PhysRevX.13.041022}. 

\section{Discussion \label{sec:disc}}
In this work, we have introduced an efficient method to incorporate spin locking into gate realizations with superconducting qubits and showed that our methods could be used with different experimentally feasible parameters. As we explained in the introduction, our method is modular, so the experimentalist could choose which data qubit to spin lock, depending on experimental considerations and the $T_\phi$ of the qubit. Although our method could be used with fixed transmons as well as tunable transmons, the latter are usually more $T_\phi$-limited and may therefore show larger improvements in gate fidelities. 
Additionally, to understand the effect of errors on the surface code, we suggested using a finer measure than the infidelity to represent the noise channel accurately. Specifically, by using the general Pauli channel, we designed a method that exploits the asymmetry between the phase noise on the protected data qubit and the unprotected ancilla qubit.

An important question that could be asked is how classical cross-talk due to the continuous microwave drive would affect neighboring qubits. Since we suggest spin locking only the data qubit in a surface code architecture, the cross-talk due to the drive would only shift the frequency of the ancilla qubit, which is dispersively coupled to the driven data qubit. Moreover, as we showed in previous sections, since both the data and the ancilla qubits could be tunable (where the data is protected, and the ancilla suffers from phase noises), the added degree of freedom would help reduce even further any cross-talk effects.

While we have focused on a specific two-qubit gate \cite{PhysRevLett.127.080505}, our method is general and could potentially be incorporated into a variety of two-qubit transmon gates. Furthermore, because changing various control parameters in time introduces time-dependent Stark shifts, a potential improvement that we leave for future research is to change the drive frequency (or control the flux for tunable transmons) during one-qubit gates, two-qubit gates, and ramp pulses, see Appendix~\ref{appendix:time-dependent transmon frequency}. Additionally, the built-in ability to detect leakage reduces leakage errors to erasures, which could remove the need to use a leakage reduction unit in the surface code. Moreover, we can utilize the fact that the Hamiltonians [Eq.\ (\ref{eq:2q_Hamiltonian})] of overlapping qubit pairs commute to perform the corresponding gates simultaneously.    

We expect that the projected efficiency of spin locking [Fig.~\ref{fig:t_solution}(c)] could push the coherence time to the $2T_1$ limit, thus making the fidelities $T_1$-limited only. This could motivate the development of superconducting qubits with long $T_1$ and short $T_\phi$. Additionally, this will favor the utilization of tunable transmons over fixed ones in future architectures, and make the frequency allocation task simpler.

\begin{acknowledgments}
We acknowledge fruitful discussions with P.~Bienias, H.~Levine, A.~Rotem and Y.~Vaknin. We are also grateful to the AWS Center for Quantum Computing for supporting this work.
\end{acknowledgments}

\appendix
\section{Diagonalization of the spin-locked qubit Hamiltonian}
\label{appendix:Eigenenergies and eigenstates of the SL qubit}
In this section, we present additional details regarding the eigenstates spin-locked transmon. Recall that we label the eigenvalues and eigenstates of Eq.~(\ref{eq:RWA_Hamiltonian}) in the main text as $\tilde E_i$ and $\ket{\tilde i}$, respectively, with $i = 0, 1, \dots$. The eigenstates $\ket{\tilde i}$ are the columns of the change-of-basis unitary matrix $U$, and the matrix of the annihilation operator in the dressed basis can be written as $U^\dagger a U$, where $a$ is the matrix of the annihilation operator in the original Fock basis. 

We find that the first 5 rows of the first 5 columns of $U$, to second order in $\epsilon$, are
\begin{equation} 
\begin{gathered}
U = \\ \left(\begin{array}{ccccc}
\frac{1}{\sqrt{2}} & \frac{1}{\sqrt{2}} & \frac{\epsilon^{2}}{2\sqrt{2}} & 0 & 0\\
\frac{1}{\sqrt{2}}-\frac{\epsilon^{2}}{4\sqrt{2}} & -\frac{1}{\sqrt{2}}+\frac{\epsilon^{2}}{4\sqrt{2}} & -\frac{\text{\ensuremath{\epsilon}}}{\sqrt{2}} & \frac{\epsilon^{2}}{4\sqrt{6}} & 0\\
\frac{\text{\ensuremath{\epsilon}}}{2}-\frac{\epsilon^{2}}{4} & -\frac{\epsilon}{2}-\frac{\epsilon^{2}}{4} & 1-\frac{11\epsilon^{2}}{32} & -\frac{\sqrt{3}\epsilon}{4} & \frac{\epsilon^{2}}{10\sqrt{3}}\\
\frac{\epsilon^{2}}{4\sqrt{3}} & -\frac{\epsilon^{2}}{4\sqrt{3}} & \frac{\sqrt{3}\text{\ensuremath{\epsilon}}}{4} & 1-\frac{43\epsilon^{2}}{288} & -\frac{\epsilon}{3}\\
0 & 0 & \frac{\sqrt{3}\epsilon^{2}}{20} & \frac{\text{\ensuremath{\epsilon}}}{3} & 1 + \mathcal{O}(\epsilon^2)
\end{array}\right) \\ + \mathcal{O}(\epsilon^3).
\end{gathered}
\end{equation}

We further find that the first 5 rows of the first 5 columns of $U^{\dagger}aU$, to second order in $\epsilon$, are
\begin{equation}
\begin{gathered}
U^{\dagger}aU= \\ \left(\begin{array}{ccccc}
\frac{1+\epsilon}{2}-\frac{3\text{\ensuremath{\epsilon}}^{2}}{8} & \frac{-1-\epsilon}{2}-\frac{\text{\ensuremath{\epsilon}}^{2}}{8} & \frac{2-\epsilon}{2}-\frac{7\text{\ensuremath{\epsilon}}^{2}}{32} & \frac{6\text{\ensuremath{\epsilon}} - 5\text{\ensuremath{\epsilon}}^{2}}{8\sqrt{3}} & \frac{\text{\ensuremath{\epsilon}}^{2}}{10\sqrt{3}}\\
\frac{1-\epsilon}{2}+\frac{\text{\ensuremath{\epsilon}}^{2}}{8} & \frac{-1+\epsilon}{2}+\frac{3\text{\ensuremath{\epsilon}}^{2}}{8} & \frac{-2-\epsilon}{2}+\frac{7\text{\ensuremath{\epsilon}}^{2}}{32} & \frac{-5\text{\ensuremath{\epsilon}}^{2} - 6\text{\ensuremath{\epsilon}}}{8\sqrt{3}} & \frac{-\text{\ensuremath{\epsilon}}^{2}}{10\sqrt{3}}\\
0 & 0 & -\frac{\epsilon}{4} & \frac{144 - 11\text{\ensuremath{\epsilon}}^{2}}{48\sqrt{3}} & \frac{\text{\ensuremath{\epsilon}}}{2\sqrt{3}}\\
0 & 0 & -\frac{\text{\ensuremath{\epsilon}}^{2}}{80\sqrt{3}} & -\frac{\text{\ensuremath{\epsilon}}}{12} & \frac{288-23\epsilon^2}{144}\\
0 & 0 & 0 & -\frac{11\text{\ensuremath{\epsilon}}^{2}}{90} & -\frac{2\epsilon}{3}
\end{array}\right) \\ + \mathcal{O}(\epsilon^3).\label{eq:creation_rotated}
\end{gathered}
\end{equation}

As we mentioned in the main text, the main implication of Eq.~(\ref{eq:creation_rotated}) is the absence of selection rules. We see that the computational levels (i.e., the first two levels) are coupled to the rest of the states. This means that gates that use a transverse coupling, such as one-qubit gates that couple microwave drive to the charge operator, or two-qubits gates that use a charge-charge coupling, could be more prone to leakage. Thus, more care is needed to eliminate leakage during such gates.

\section{Transverse coupling}
\label{appendix:Transverse coupling}

In this section, we show how the clock condition in Eq.~(\ref{eq:insensativity}) in the main text also implies protection from population fluctuations of a system that is dispersively coupled to the spin-locked qubit, as we mention in the main text. Assume
there is a harmonic system or a TLS (two-level system) that is transversely weakly coupled to
a transmon (i.e.~via charge-charge interaction). Using the Kerr Hamiltonian for the transmon, within the rotating-wave approximation, we get
\begin{equation}
H=\omega a^{\dagger}a-\frac{\eta}{2}a^{\dagger2}a^{2}+\frac{\Omega}{2}(ae^{i\omega_{d}t}+h.c)+g(a^{\dagger}b+h.c)+\omega_{b}b^{\dagger}b,
\end{equation}
where $b$ is the creation operator for the parasitic system, either a harmonic system or a TLS, $\omega_b$ is the frequency of the parasitic system, and $g$ is the coupling strength between the systems.

\begin{figure}[htp]
\centering
\includegraphics[width=8.6cm]{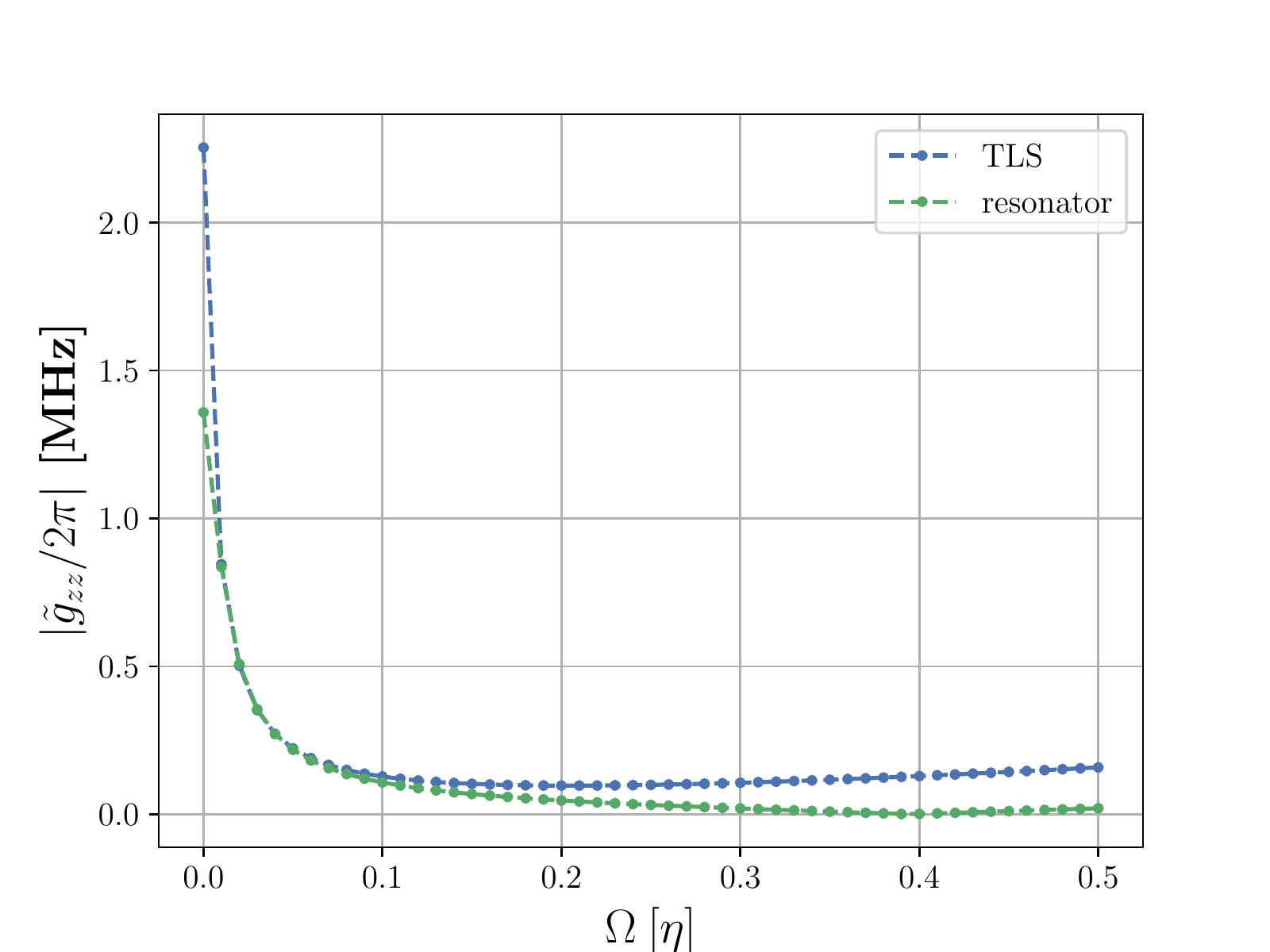}

\caption{$|\frac{\tilde{g}_{zz}}{2\pi}|$ for the spin-locked qubit coupled to a TLS (blue) or a resonator (green). The coupling is $\frac{g}{2\pi}=70$ MHz for the resonator and $\frac{g}{2\pi}=50$ MHz for the TLS. The qubit's frequency is $\frac{\omega}{2\pi}=5$ GHz and the second system's frequency is $\frac{\omega_b}{2\pi}=6.1$ GHz. The qubit's anharmonicity is  $\frac{\eta}{2\pi}=300$ MHz. The drive's detuning is set to be $\delta=\delta(\Omega)+\frac{g^2}{w-\omega_b}$. To make sure that the contribution from higher levels of the spin-locked qubit is taken correctly into account, the simulation was done using the full cosine potential (see Sec.~\ref{section: Full model - cosine potential}). The numerical diagonalization presented here, as well as all of the numerical simulations presented in the main text and in the supplemental material, were done using QuTiP \cite{JOHANSSON20121760_revised}.} 

\label{fig:tls_resonator}
\end{figure}

Moving to a rotating frame with respect to the drive frequency, we get
\begin{equation}
H_{I}=-\delta a^{\dagger}a-\frac{\eta}{2}a^{\dagger2}a^{2}+\frac{\Omega}{2}(a+a^{\dagger})+g(a^{\dagger}b+h.c)-\Delta b^{\dagger}b,
\end{equation}
where $\delta = \omega_d - \omega$ and $\Delta=\omega_{d}-\omega_{b}$. Without spin locking, in the limit of large $|\omega_b - \omega|$, the effective interaction would be $g_{zz}a^\dagger a b^\dagger b$. Therefore, to lowest non-vanishing order in $\frac{g}{\Delta}$ and $\frac{\Omega}{\eta}$, spin locking also deals with this interaction. A more detailed analysis requires moving to the dressed basis of
the driven transmon, so that the Hamiltonian becomes
\begin{equation}
H_{I}=\sum_{\tilde{n}}\tilde{E}_{n}|\tilde{n}\rangle\langle\tilde{n}|-\Delta b^{\dagger}b+g(a^{\dagger}b+h.c),
\label{eq:coupled_system}
\end{equation}
where, as in the main text, $\{\tilde{E}_i\}$ ($\{|\tilde{i}\rangle\}$) are the dressed energies (states). Thus, it is apparent that in the limit $|\tilde{E}_n-\tilde{E}_m-\Delta| \gg g$, where $|\tilde{m}\rangle$ and $|\tilde{n}\rangle$ are the dressed states which are coupled to the dressed states in the dressed computational subspace, we could approximate the transverse coupling as a cross-Kerr interaction $\tilde{g}_{zz} (|\tilde{0}\rangle\langle\tilde{0}|-|\tilde{1}\rangle\langle\tilde{1}|) b^{\dagger}b$, where $\tilde{g}_{zz}\ll g_{zz}$. Figure \ref{fig:tls_resonator} shows the exact $ZZ$ strength for two types of systems, i.e.\ $\tilde{g}_{zz}$ for the driven qubit coupled to a resonator and coupled to a TLS. The point where the drive reaches zero shows the bare ${g}_{zz}$, i.e., $\tilde{g}_{zz}(\Omega=0)=g_{zz}$. We see a significant reduction of the $ZZ$ interaction strength in the presence of spin locking, i.e.\ when $\Omega$ is away from zero. We note that due to the Stark shift that is induced by the second system, the qubit's frequency is effectively $\omega\rightarrow \omega+\frac{g^2}{w-\omega_b}$. To account for that, we change the drive's detuning accordingly: $\delta=\delta(\Omega)+\frac{g^2}{w-\omega_b}$, where $\delta(\Omega)$ is the function described in the main text, and $\omega$ is the original qubit frequency.

\subsection{Transverse coupling with a harmonic system}

Let us now analyze transverse coupling to a harmonic oscillator in more detail. Starting from Eq.~(\ref{eq:coupled_system}) and 
using second-order perturbation theory, we find that the energy $E_{\tilde{j}k}$ of the $|\tilde{j}\rangle$ dressed basis state and Fock state $|k\rangle$ of the resonator is
\begin{align}E_{\tilde{j}k} 
 & = g^{2}\sum_{\tilde{n}}\frac{|\alpha_{\tilde{j},\tilde{n}}|^{2}}{(\tilde{E}_{j}-\tilde{E}_{n})-\Delta}k+\frac{|\alpha_{\tilde{n},\tilde{j}}|^{2}}{(\tilde{E}_{j}-\tilde{E}_{n})+\Delta}(k+1), 
\end{align}
where we defined $\alpha_{\tilde{n},\tilde{j}}=\langle\tilde{n}|a|\tilde{j}\rangle$.
Thus, $\tilde{g}_{zz}$ is the part that depends
linearly on the Fock state $k$:
\begin{align}
&
\tilde{g}_{zz}=\frac{g^{2}}{2}\sum_{\tilde{n}}\frac{|\alpha_{\tilde{0},\tilde{n}}|^{2}}{(\tilde{E}_{0}-\tilde{E}_{n})-\Delta}+\frac{|\alpha_{\tilde{n},\tilde{0}}|^{2}}{(\tilde{E}_{0}-\tilde{E}_{n})+\Delta} \\ &-\frac{|\alpha_{\tilde{1},\tilde{n}}|^{2}}{(\tilde{E}_{1}-\tilde{E}_{n})-\Delta}-\frac{|\alpha_{\tilde{n},\tilde{1}}|^{2}}{(\tilde{E}_{1}-\tilde{E}_{n})+\Delta}\nonumber. 
\end{align}

Further assuming that $\Delta$
is the largest energy scale in the system, i.e.~$|\Delta|\gg|\tilde{E}_{0}-\tilde{E}_{n}|,|\tilde{E}_{1}-\tilde{E}_{n}|$, we obtain, to first order in $1/\Delta$: 
\begin{align}
\tilde{g}_{zz} & \approx\frac{g^{2}}{2\Delta}\sum_{\tilde{n}}|\alpha_{\tilde{0},\tilde{n}}|^{2}-|\alpha_{\tilde{1},\tilde{n}}|^{2}-|\alpha_{\tilde{n},\tilde{0}}|^{2}+|\alpha_{\tilde{n},\tilde{1}}|^{2}.
\end{align}

Let us now recall the clock condition from Eq.~(\ref{eq:insensativity}):
\begin{align}
 & \langle\tilde{0}|a^{\dagger}a|\tilde{0}\rangle=\langle\tilde{1}|a^{\dagger}a|\tilde{1}\rangle\Rightarrow\\
 & \sum_{\tilde{n}}\alpha_{\tilde{n}\tilde{0}}^{*}\alpha_{\tilde{n}\tilde{0}}=\sum_{\tilde{n}}\alpha_{\tilde{n}\tilde{0}}^{*}\alpha_{\tilde{n}\tilde{0}}\Rightarrow\nonumber \\
 & \sum_{\tilde{n}}|\alpha_{\tilde{n}\tilde{0}}|^{2}-|\alpha_{\tilde{n}\tilde{1}}|^{2}=0.\nonumber 
\end{align}
Furthermore, since the dressed states are normalized, we have
\begin{align}
 & \langle\tilde{0}|a^{\dagger}a|\tilde{0}\rangle=\langle\tilde{1}|a^{\dagger}a|\tilde{1}\rangle\Rightarrow\\
 & \langle\tilde{0}|aa^{\dagger}|\tilde{0}\rangle=\langle\tilde{1}|aa^{\dagger}|\tilde{1}\rangle\Rightarrow\nonumber \\
 & \sum_{\tilde{n}}|\alpha_{\tilde{0}\tilde{n}}|^{2}-|\alpha_{\tilde{1}\tilde{n}}|^{2}=0.\nonumber 
\end{align}
Therefore, we see that the clock condition from Eq.~(\ref{eq:insensativity}) in the main text implies that $\tilde{g}_{zz}=0$ to first order in $1/\Delta$.
However, the converse isn't true. Moreover, it can be easily seen that higher-order terms in $1/\Delta$ do not generally vanish as a result of Eq.~(\ref{eq:insensativity}), meaning that the
condition $|\Delta|\gg|\tilde{E}_{0}-\tilde{E}_{n}|,|\tilde{E}_{1}-\tilde{E}_{n}|$ is crucial for ensuring that Eq.~(\ref{eq:insensativity}) implies the vanishing of $\tilde{g}_{zz}$.

\subsection{Transverse coupling with a two-level system}

Now we assume the transmon is coupled to a two-level system (TLS). The
Hamiltonian is
\begin{equation}
H =\omega a^{\dagger}a-\frac{\eta}{2}a^{\dagger2}a^{2}+\frac{\Omega}{2}(ae^{i\omega_{d}t}+h.c) g(a^{\dagger}\sigma_{-}+h.c)+\frac{\omega_{\sigma}}{2}\sigma_{z}.
\end{equation}
Defining $\Delta=\omega_d-\omega_{\sigma}$ and using the same procedure,  we find the energies $E_{\tilde{j}\uparrow}$ and $E_{\tilde{j}\downarrow}$ of the $|\tilde{j}\rangle$ dressed basis state combined with the $\ket{\uparrow}$ and $\ket{\downarrow}$ states, respectively, of the two-level system:
\begin{align}
 & E_{\tilde{j}\downarrow}=g^{2}\sum_{\tilde{n}}\frac{|\alpha_{\tilde{n},\tilde{j}}|^{2}}{(\tilde{E}_{j}-\tilde{E}_{n})+\Delta},\\
 & E_{\tilde{j}\uparrow}=g^{2}\sum_{\tilde{n}}\frac{|\alpha_{\tilde{j},\tilde{n}}|^{2}}{(\tilde{E}_{j}-\tilde{E}_{n})-\Delta}.
\end{align}
Therefore, the $ZZ$ interaction strength is
\begin{align}
 \tilde{g}_{zz}
 & = \frac{g^{2}}{4}\sum_{\tilde{n}}\frac{|\alpha_{\tilde{n},\tilde{0}}|^{2}}{(\tilde{E}_{0}-\tilde{E}_{n})-\Delta}-\frac{|\alpha_{\tilde{n},\tilde{1}}|^{2}}{(\tilde{E}_{1}-\tilde{E}_{n})-\Delta} \\ & -\frac{|\alpha_{\tilde{0},\tilde{n}}|^{2}}{(\tilde{E}_{0}-\tilde{E}_{n})+\Delta}+\frac{|\alpha_{\tilde{1},\tilde{n}}|^{2}}{(\tilde{E}_{1}-\tilde{E}_{n})+\Delta}. \nonumber
\end{align}
Again, assuming $|\Delta|\gg|\tilde{E}_{0}-\tilde{E}_{n}|,|\tilde{E}_{1}-\tilde{E}_{n}|$, we find, to first order in $1/\Delta$,
\begin{equation}
\tilde{g}_{zz}\approx\frac{g^{2}}{4\Delta}\sum_{n}-|\alpha_{\tilde{n},\tilde{0}}|^{2}+|\alpha_{\tilde{n},\tilde{1}}|^{2}-|\alpha_{\tilde{0},\tilde{n}}|^{2}+|\alpha_{\tilde{1},\tilde{n}}|^{2}.
\end{equation}
Thus, the clock condition from Eq.~(\ref{eq:insensativity}) in the main text again implies the vanishing of $g_{zz}$ to first order in $1/\Delta$.

\section{one-qubit gate: additional information}
\label{appendix: one-qubit gate}

\begin{figure}[t]
\centering
\includegraphics[width=8.6cm]{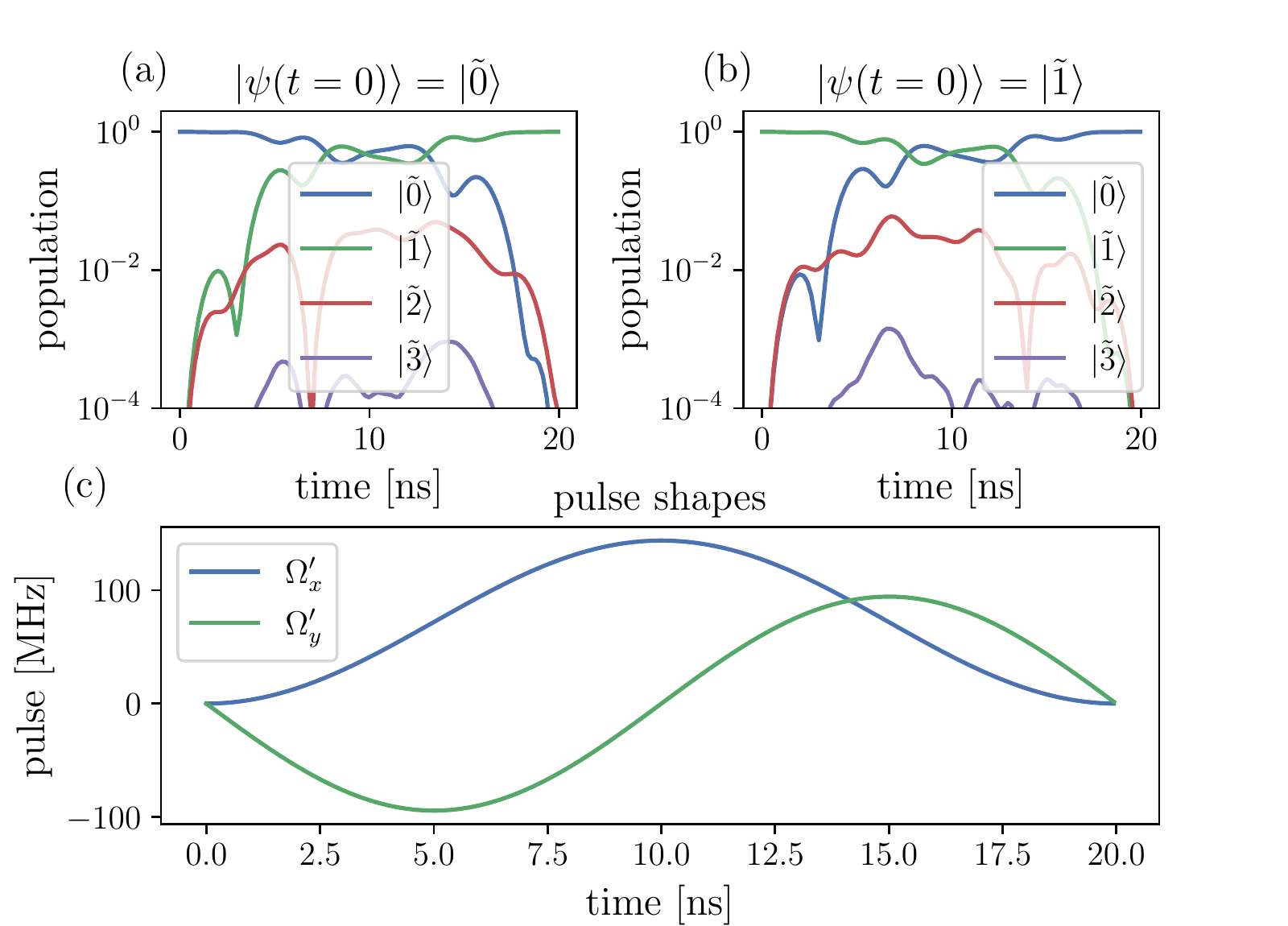}

\caption{$X$ gate for the dressed qubit with $\Omega = 0.3 \eta$.  (a) and (b) show populations of the dressed (Floquet) eigenstates. Although the points were taken only at integer multiples of $\frac{2\pi}{\omega_d}$, the graphs show smooth curves for readability. The initial state is (a) $|\tilde{0}\rangle$ or (b)  $|\tilde{1}\rangle$. (c) shows the pulse shape, which is explained in Eq.~(\ref{eq:1_q_gate_pulse_shape}).
The parameters for this simulation are $\frac{\omega}{2\pi}=5$ GHz, $\frac{\eta}{2\pi}=300$ MHz, $A=-3.2\pi$, $l_{D}=7.88$, and $\frac{\omega_{d}'}{2\pi}=5157.761$ MHz.
$\delta$ is chosen to be on the $\delta(\Omega)$ curve. The gate time is $T_g=20$ ns, and the average infidelity is $5.53\cdot10^{-5}$.}
\label{fig:1_q_gate_0.3}
\end{figure}

\begin{figure}[t]
\centering
\includegraphics[width=8.6cm]{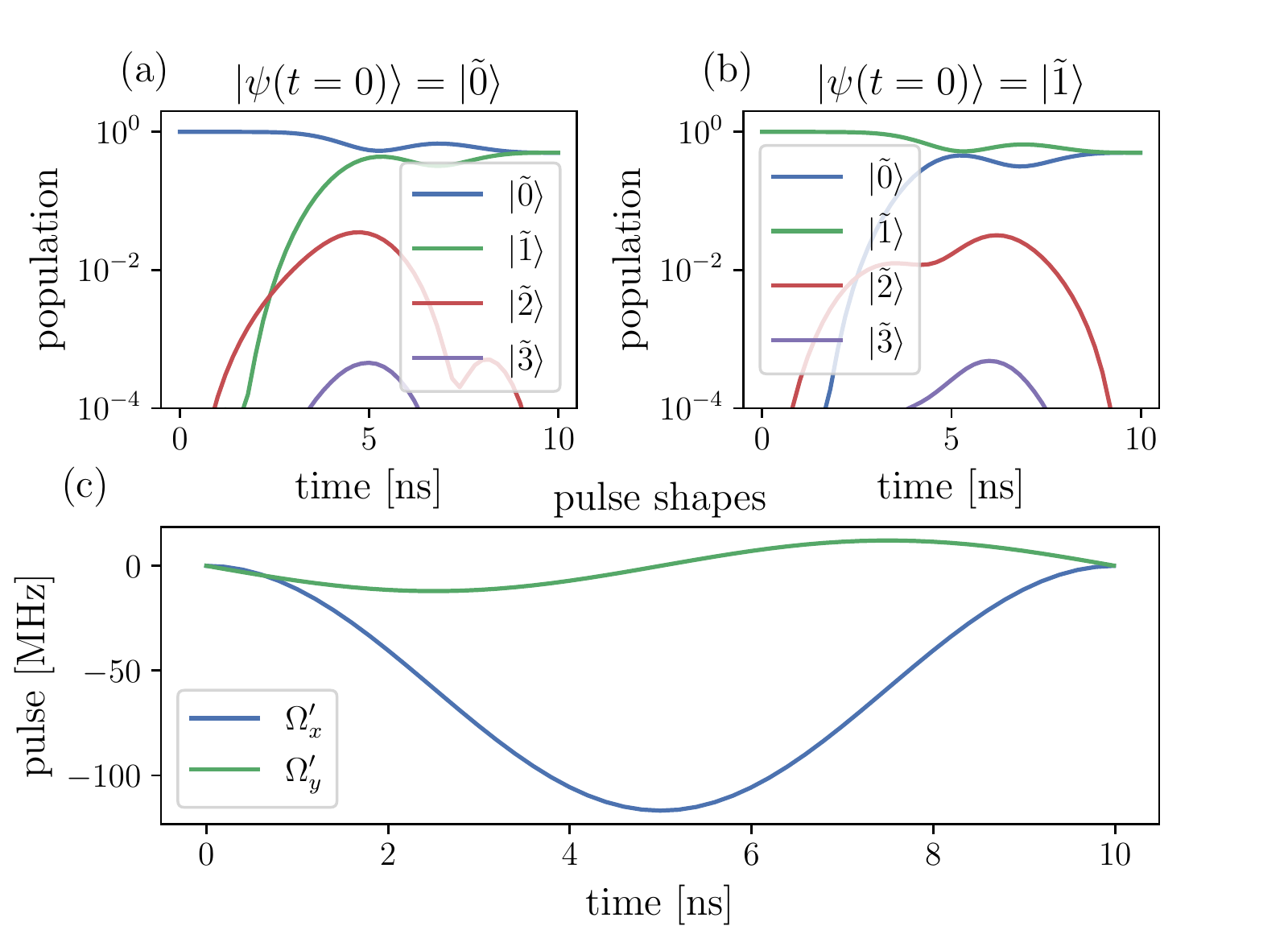}

\caption{$X_{\frac{\pi}{4}}$ gate for the dressed qubit with $\Omega = 0.2 \eta$.  (a) and (b) show populations of the dressed (Floquet) eigenstates. Although the points were taken only at integer multiples of $\frac{2\pi}{\omega_d}$, the graphs show smooth curves for readability. The initial state is (a) $|\tilde{0}\rangle$ or (b)  $|\tilde{1}\rangle$. (c) shows the pulse shape, which is explained in Eq.~(\ref{eq:1_q_gate_pulse_shape}).
The parameters for this simulation are $\frac{\omega}{2\pi}=5$ GHz, $\frac{\eta}{2\pi}=300$ MHz, $A=-1.29\pi$, $l_{D}=-0.62$, and $\frac{\omega_{d}'}{2\pi}=5080.096$ MHz.
$\delta$ is chosen to be on the $\delta(\Omega)$ curve. The gate time $T_g=10$ ns, and the average infidelity is $5.02\cdot10^{-6}$. The gate was optimized for $Z$ rotations in the dressed basis, i.e., the gate is $X_{\frac{\pi}{4}}(\theta_1, \theta_2)=e^{i\theta_1 \tilde{\sigma}_z}X_{\frac{\pi}{4}}e^{i\theta_2 \tilde{\sigma}_z}$ for some $\theta_1, \theta_2$, where $\tilde{\sigma}_z$ is the Pauli $Z$ matrix in the dressed basis. The difference between $X_{\frac{\pi}{4}}$ and $X_{\frac{\pi}{4}}(\theta_1, \theta_2)$ can be taken into account in software during circuit compilation \cite{Krantz_2019_revised}.}
\label{fig:hadamard}
\end{figure}

In this section, we present the simulation results for the $X$ gate with $\Omega=0.3\eta$ and the $e^{i\sigma_x\frac{\pi}{4}}$ gate with $\Omega=0.2\eta$. The $X$ gate simulation and details can be seen in Fig.~\ref{fig:1_q_gate_0.3}, and similarly for the $e^{i\sigma_x\frac{\pi}{4}}$ gate [Fig.~\ref{fig:hadamard}]. Both of the gates are fast (20 ns and 10 ns correspondingly) and with low average infidelity ($5.53\cdot10^{-5}$ and $5.02\cdot 10^{-6}$ correspondingly).

\section{two-qubit gate}
\label{appendix: two-qubit gate}

In this section, we provide more details regarding the two-qubit gate and show the evolution of the populations during the gate.

\subsection{Sensitivity measure in a hybridized system}
\label{appendix: two-qubit gate - sensativity}

To understand how the gate performs under phase noise, we want to track the sensitivity measure, as shown in Fig.~\ref {fig:g_zz for coupler and pulse shape}. As we explained in Appendix \ref{appendix:Transverse coupling}, due to Stark shifts from the other system (in this case, $q_2$ and the coupler) the detuning of the drive $\delta$ should be shifted accordingly. Moreover, due to hybridization, higher-order corrections to the eigenstates would create a difference between the sensitivity measure in different configurations of the coupled systems. As an example, denote the hybridized state by $|\tilde{i}jk\rangle$, where $q_1$ is in state $\ket{\tilde{i}}$, $q_2$ in state $\ket{j}$, and the coupler in state $\ket{k}$. Due to hybridization, the sensitivity measure where $q_2$ and the coupler are in the ground state, $|\langle\tilde{0}00 |a^\dagger a|\tilde{0}00 \rangle - \langle\tilde{1}00 |a^\dagger a|\tilde{1}00 \rangle|$, will not be equal to the sensitivity measure where $q_2$ is in the first excited state and the coupler is in its ground state,   $|\langle\tilde{0}10 |a^\dagger a|\tilde{0}10 \rangle - \langle\tilde{1}10 |a^\dagger a|\tilde{1}10 \rangle|$.   

We stress that we want the driven qubit to be protected from arbitrary sources of phase noise, regardless of the configuration of the rest of the system. Since most of the contribution of the other coupled systems to the driven qubit is described by the Stark shift (which can be taken into account when choosing $\delta(\Omega)$),  
the different sensitivities in different configurations would give similar results in terms of gate fidelity. Additionally, in an experimental context, in order to minimize the phase noise sensitivity originating from different noise mechanisms, a more experimentally accessible metric would be to maximize $T_{2,\rho}$ over the choice of $\delta$.

In light of this discussion, we decided to choose a symmetric sensitivity measure between $q_2$ computational states:  
\begin{align}
    \frac{1}{2}\{|\langle\tilde{0}00 |a^\dagger a|\tilde{0}00 \rangle - \langle\tilde{1}00 |a^\dagger a|\tilde{1}00 \rangle| + \nonumber\\
    |\langle\tilde{0}10 |a^\dagger a|\tilde{0}10 \rangle - \langle\tilde{1}10 |a^\dagger a|\tilde{1}10 \rangle|\}.
    \label{eq:ins_hybrid}
\end{align}    
As explained above, this choice is not unique and gives similar results to the cases when we only take into account the Stark shift  
or choose only one of the computational states of $q_2$ to define the sensitivity measure. We leave for future research the quest to find the optimal sensitivity measure for hybridized systems; such a measure will depend on how the systems coupled to the driven qubit are used during the computation.

 We point out that, during the idling stage (i.e.\ when we are not applying a two-qubit gate), we would like to have both minimal $g_{zz}$ and minimal sensitivity. Thus, we choose the drive frequency $w_d$ and the idling coupler frequency $\omega_{c}(0)$ to minimize both quantities simultaneously. In particular, we use the steps in the following iterative approach to find $\omega_d$ and $\omega_c(0)$. (1) We choose $w_d$ to be on the $\delta(\Omega)$ curve, including the Stark shift. (2) We find $w_c$ that minimizes $\tilde{g}_{zz}$. (3) We corrected $w_d$ to minimize the sensitivity in Eq.~(\ref{eq:ins_hybrid}). (4) Returned to step 2 until convergence.  

\subsection{Details regarding gates with different parameters}
\label{appendix: two-qubit gate - realizations}

In this section, we provide details regarding the simulation of the gates in Sec.~\ref{subsec: 2q - additional gates}. %

\begin{figure}[t]
\centering
\includegraphics[width=8.6 cm]{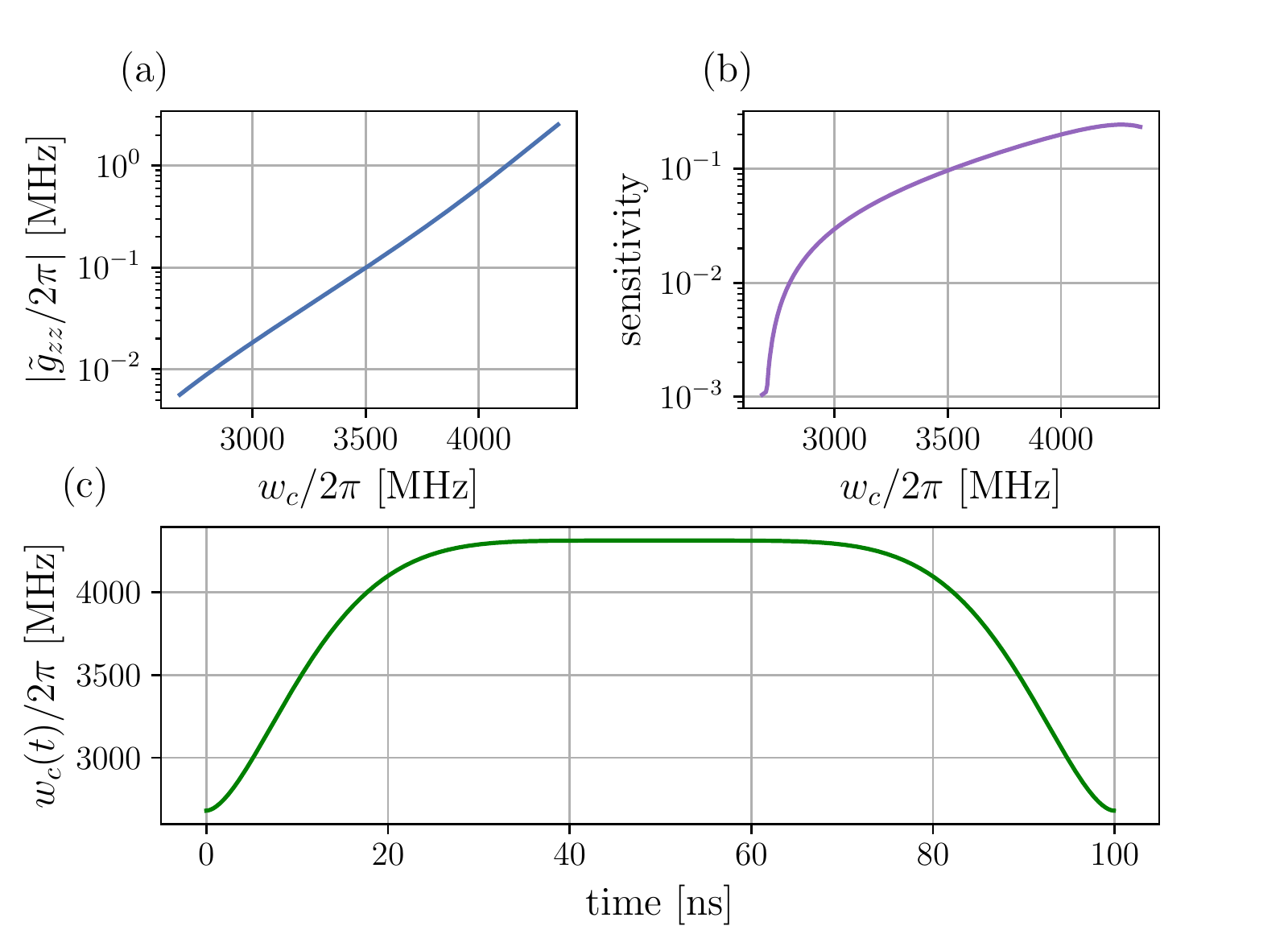}

\caption{two-qubit gate operation graphs for $\frac{g_{1c}}{2\pi}=\frac{g_{2c}}{2\pi}=150$ MHz and $\frac{g_{12}}{2\pi}=-5$ MHz. The figures are the same as in Fig.~\ref{fig:g_zz for coupler and pulse shape}.}
\label{fig:2q_params_small_couplings}
\end{figure}

Figs.~\ref{fig:2q_params_small_couplings}(a) and (b) show $\tilde{g}_{zz}$ and sensitivity curves for the case of smaller couplings than in Sec.~\ref{subsec: 2q - Pulse shape and gate parameters}. We see that, if we set the coupler at $\frac{\omega_c}{2\pi}\approx 2680$ MHz, we get small $\frac{|\tilde{g}_{zz}|}{2\pi}\approx 5$ kHz, while an operation point of $\frac{\omega_c}{2\pi}\approx 4340$ MHz would still yield relatively strong ZZ interaction of $\frac{|\tilde{g}_{zz}|}{2\pi}\approx 2.5$ MHz, as in Fig.~\ref{fig:g_zz for coupler and pulse shape}(a). We used the same pulse shape as in Eq.~(\ref{eq:2q_pulse_shape}) with $T_g=100$ ns, $\frac{\omega_s}{2\pi}=2680$ MHz, $\frac{\omega_f}{2\pi}=4312$ MHz, $T_r=49.952$ ns, $m=2.906$, and $n=1.777$. We achieved an average infidelity of $\approx 2\cdot 10^{-5}$. 

\begin{figure}[t]
\centering
\includegraphics[width=8.6 cm]{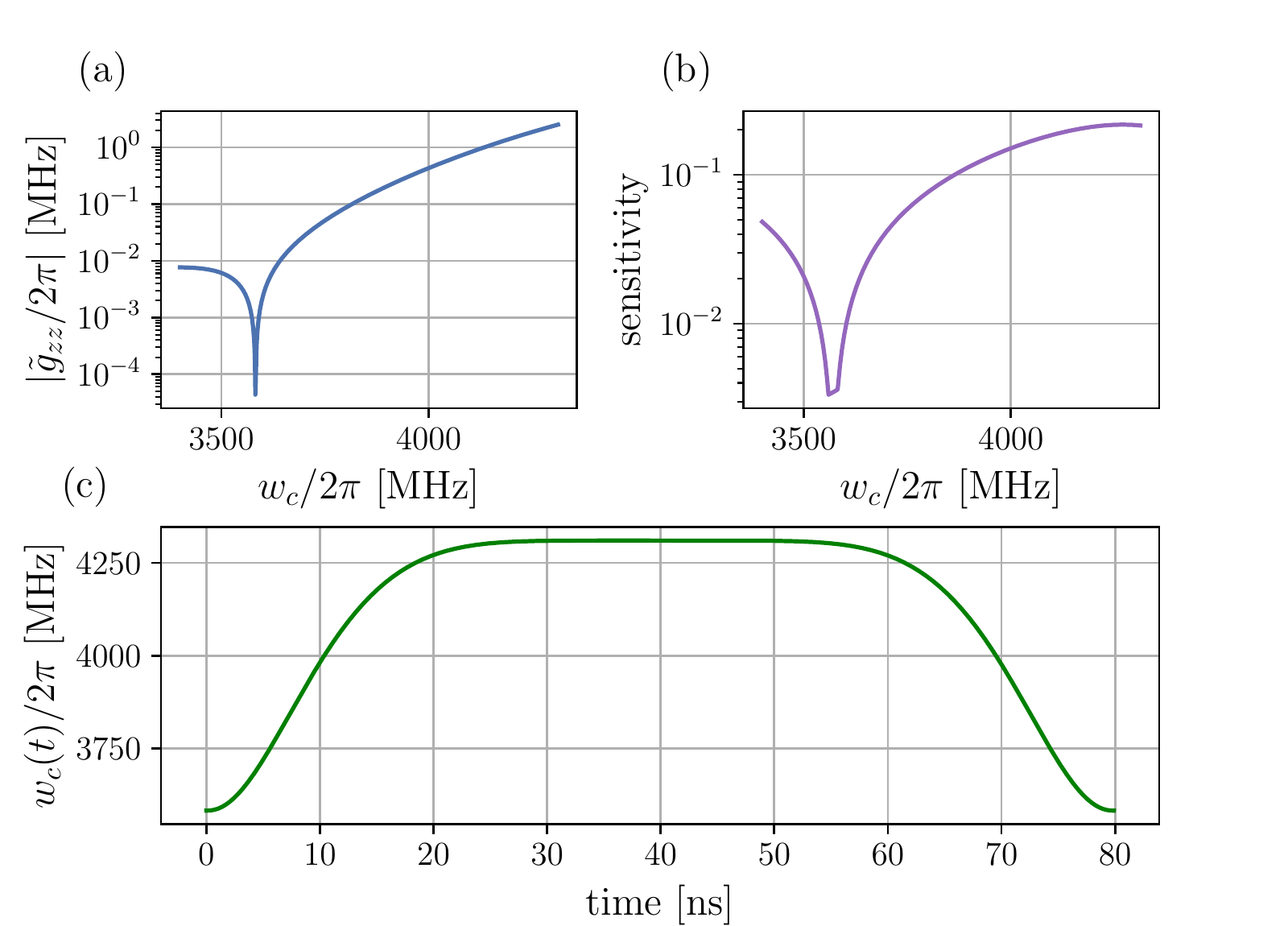}

\caption{two-qubit gate operation graphs for $T_g=80$ ns. The couplings and the description of the subfigures are the same as in Fig.~\ref{fig:g_zz for coupler and pulse shape}}
\label{fig:2q_params_fast_gate}
\end{figure}

Similarly, we simulated a faster gate ($T_g=80$ ns) with the same couplings as in Sec.~\ref{subsec: 2q - Pulse shape and gate parameters}, see Fig.~\ref{fig:2q_params_fast_gate}. The optimized parameters are $\frac{\omega_s}{2\pi}=3580$ MHz, $\frac{\omega_f}{2\pi}=4310.27$ MHz, $T_r=39.95$ ns, $m=3.044$, and $n=2.298$. We achieved an average infidelity of $\approx 3\cdot 10^{-4}$. 

\subsection{Populations graphs of the two-qubit gate}
\label{appendix: two-qubit gate - population}
\begin{figure*}[ht]
\centering
\includegraphics[width=\textwidth]{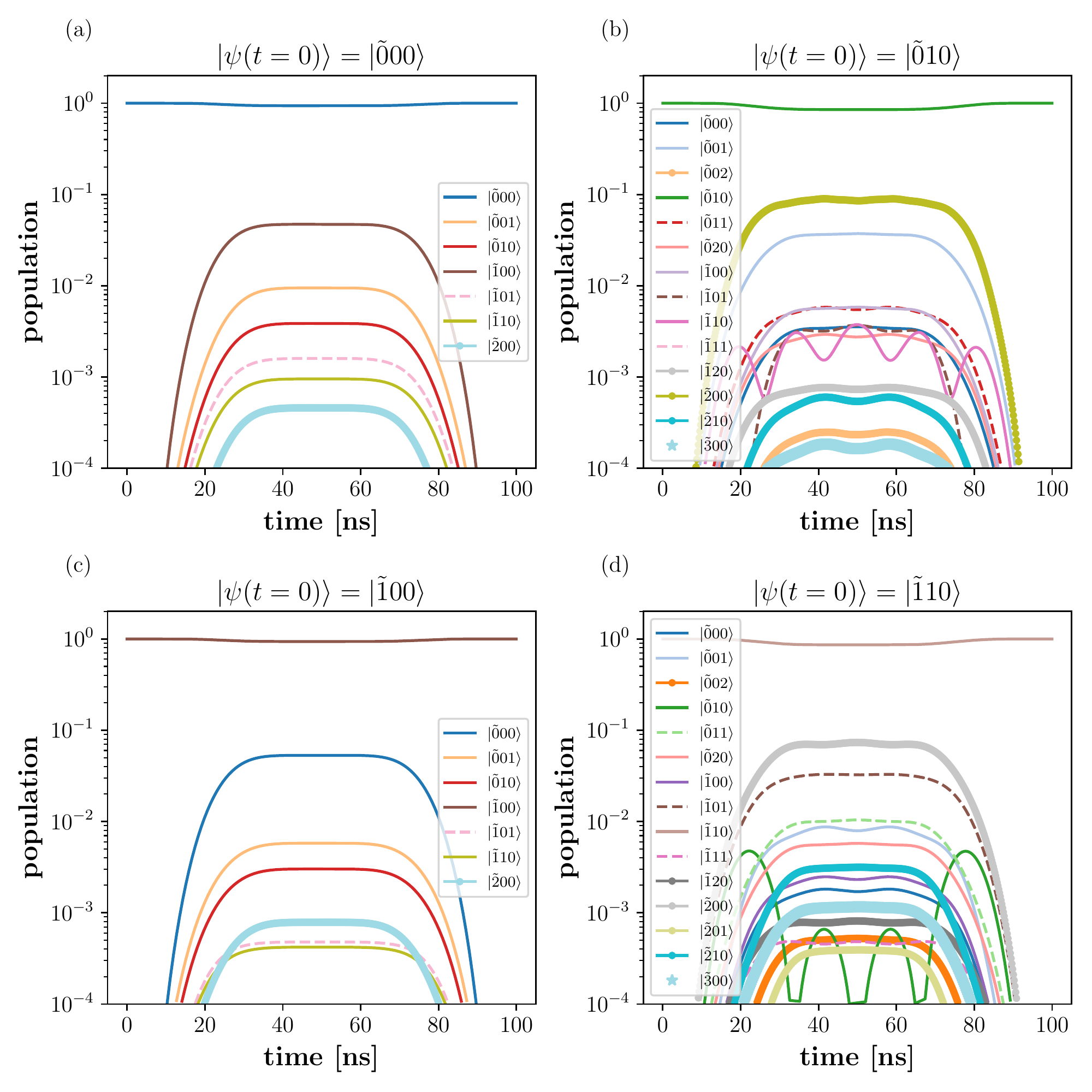}

\caption{State populations during the CZ gate presented in Sec.~\ref{subsec: 2q - Pulse shape and gate parameters} in the main text. We use the basis $|\tilde{Q1},Q2,C\rangle$, where $Q$ stands for qubit, $C$ stands for coupler, and ${Q1}$ states have a $\tilde{}$ on top to remind us that qubit 1 spin-locked. The states presented are the hybridized states of the system that correspond to tensor product states $|\tilde{Q1},Q2,C\rangle$. Although the points were calculated only at integer multiples of $\frac{2\pi}{\omega_d}$, the graphs show smooth curves for readability. The initial state is (a) $|\tilde{0}00\rangle$,  (b) $|\tilde{0}10\rangle$, (c)  $|\tilde{1}00\rangle$, or (d) $|\tilde{1}10\rangle$. The gate was optimized for single-qubit Z rotations in the dressed basis for the spin-locked qubit and in the bare basis for the non-driven qubit. The actual gate is therefore $e^{i\theta_1\tilde{\sigma}_z^1}e^{i\theta_2\sigma_z^2} CZ e^{i\theta_3\tilde{\sigma}_z^1}e^{i\theta_4\sigma_z^2}$ for some $\theta_1,\theta_2,\theta_3,\theta_4$, where $\tilde{\sigma}_z^1$ is the Pauli Z operator for the dressed qubit and $\sigma_z^2$ is the Pauli Z operator for the non-driven qubit. The extra single-qubit Z rotations can be taken into account in software during circuit compilation \cite{Krantz_2019_revised}.}  
\label{fig:2q_gate_population_main}
\end{figure*}

\begin{figure*}[t]
\centering
\includegraphics[width=\textwidth]{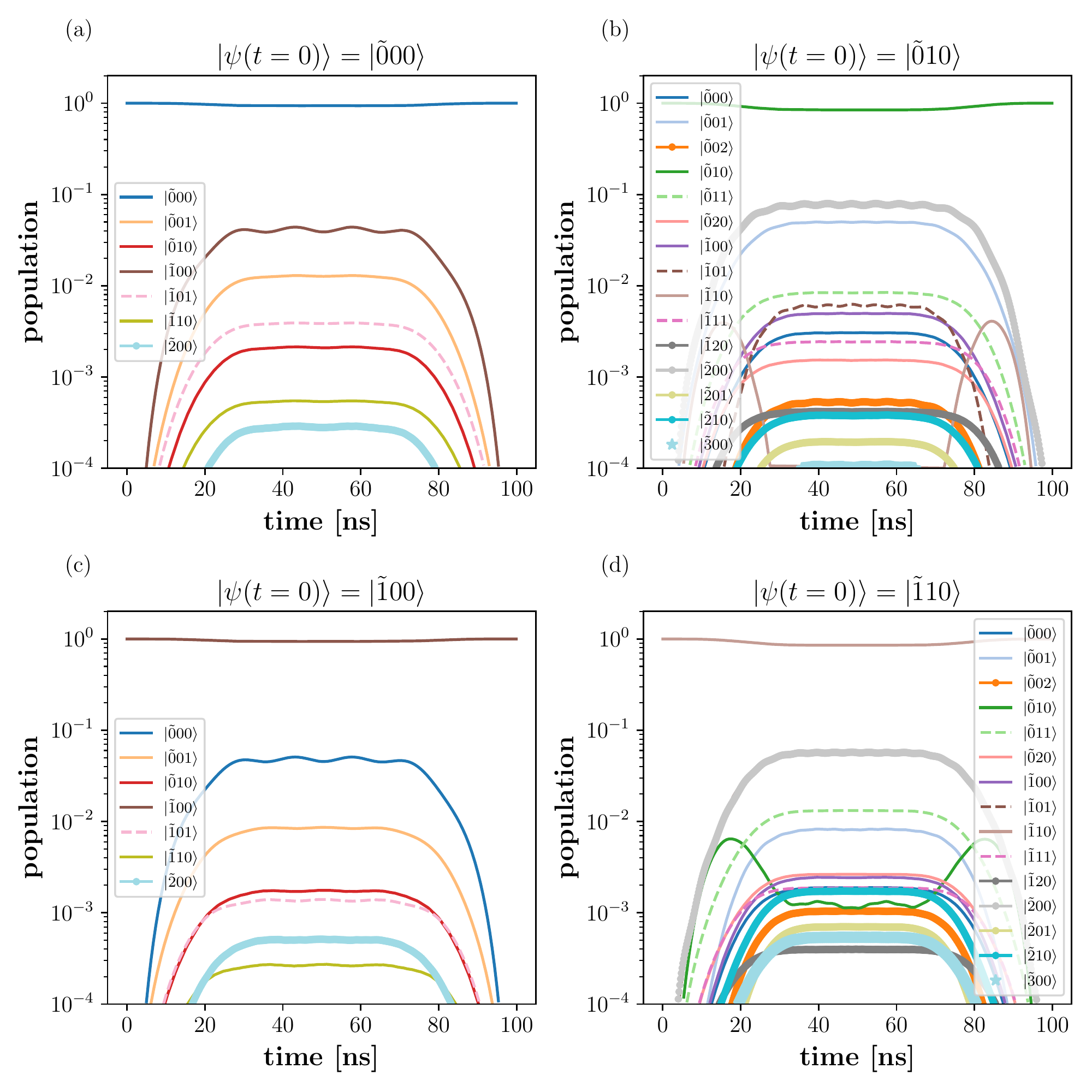}

\caption{State populations during the CZ gate for $\frac{g_{1c}}{2\pi}=\frac{g_{2c}}{2\pi}=150$ MHz and $\frac{g_{12}}{2\pi}=-5$ MHz, that is described in Appendix \ref{appendix: two-qubit gate - realizations}. The rest of the figure is similar to Fig.~\ref{fig:2q_gate_population_main}.}
\label{fig:2q_gate_population_diff}
\end{figure*}

\begin{figure*}[t]
\centering
\includegraphics[width=\textwidth]{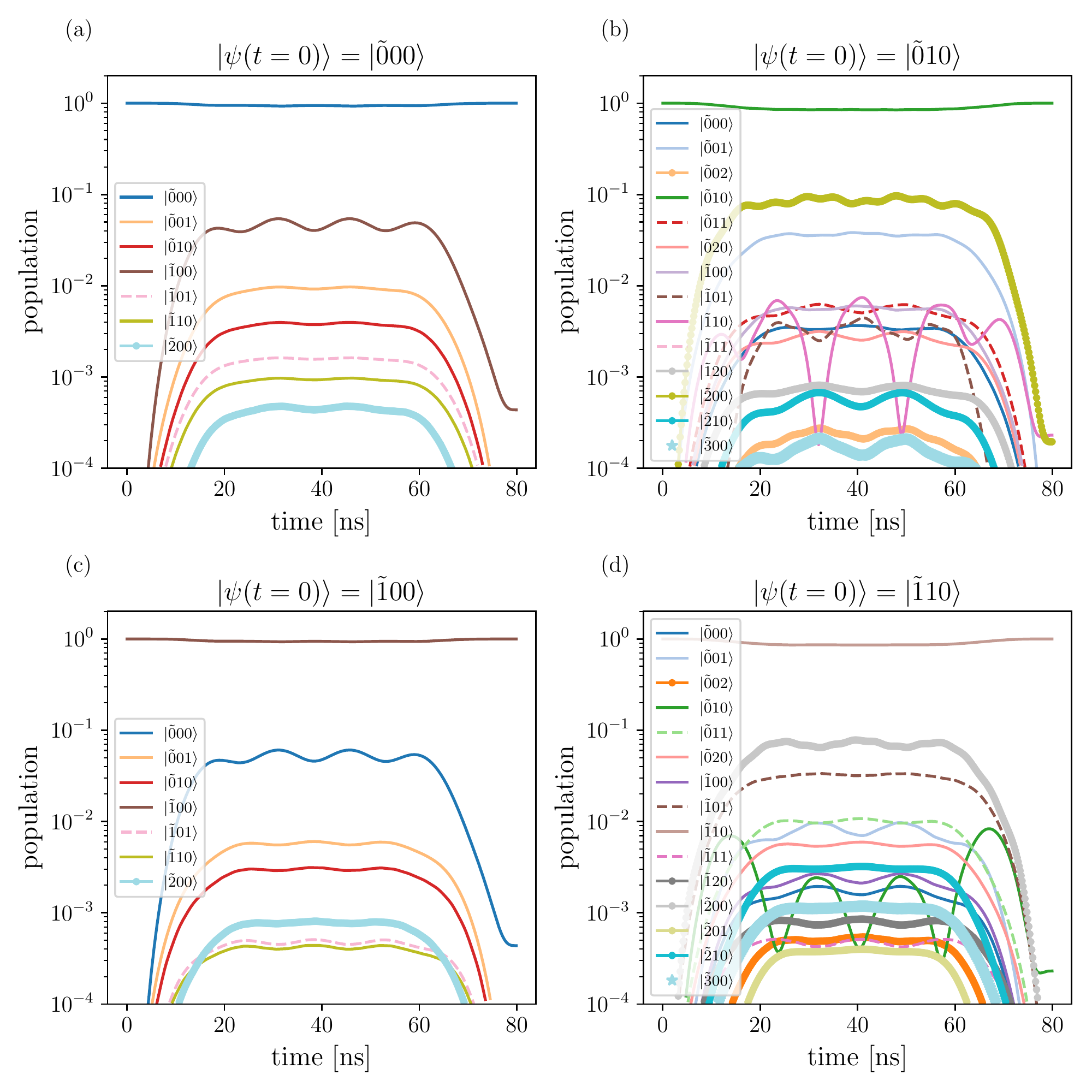}

\caption{State populations during the fast ($T_g=80$ ns) CZ gate, that is described in Appendix \ref{appendix: two-qubit gate - realizations}. The rest of the figure is similar to Fig.~\ref{fig:2q_gate_population_main}. We see that the gate suffers more from non-adiabatic errors than the longer gates.}
\label{fig:2q_gate_population_fast}
\end{figure*}

 In this section, we show the population graphs for the two-qubit gate simulations. Fig.~\ref{fig:2q_gate_population_main}, \ref{fig:2q_gate_population_diff}, \ref{fig:2q_gate_population_fast} show the population graphs for the gate described in Sec.~\ref{subsec: 2q - Pulse shape and gate parameters}, the gate with smaller couplings and the fast gate described in Appendix \ref{appendix: two-qubit gate - realizations} correspondingly.  

\section{Pauli-twirl approximation (PTA) for Z errors}
\label{appendix: PTA_for_z_errors}
In the main text, we use $P_z(T_s)=\left(\pi \frac{T_g}{T_s}\right)^2$ as the expected Z-error probability under the Pauli-twirl approximation (PTA) in the case of a bare transmon. In this section, we will derive this equation. Suppose that we have a frequency shift $\zeta=\frac{2\pi}{T_s}$, so that the Hamiltonian in a frame rotating with the qubit frequency is 
\begin{equation}
    H=\zeta a^\dagger a -\frac{\eta}{2}a^{\dagger 2} a^2,
\end{equation}
where $\eta$ is the transmon anharmonicity. Using a two-level approximation, we arrive at 
\begin{equation}
    H=\frac{\zeta}{2} \sigma_z,
\end{equation}
where $\sigma_z$ is the Z Pauli matrix. Thus, the time evolution operator is
\begin{equation}
    U=e^{-iHt}=e^{-it\frac{\zeta}{2} \sigma_z}=\cos\left(t{\zeta}/{2}\right) -i \sin\left(t{\zeta}/{2}\right)\sigma_z.
\end{equation}
Using PTA amounts to neglecting all the off-diagonal terms of the channel, which means that
\begin{equation}
    \Lambda(\rho) = U\rho U^\dagger\underset{PTA}{\rightarrow} \cos^2\left(t{\zeta}/{2}\right) \rho + \sin^2\left(t{\zeta}/{2}\right) \sigma_z \rho \sigma_z. 
\end{equation}
Setting the time $t=T_g$ to be the time of the gate and assuming $\pi\frac{T_g}{T_s}\ll1$, we arrive at
\begin{equation}
    \Lambda(\rho)_{PTA} \approx (1-(\pi{T_g}/{T_s})^2) \rho + (\pi{T_g}/{T_s})^2 \sigma_z \rho \sigma_z,
\end{equation}
which concludes the derivation.

\section{two-qubit gate with dephasing on the coupler}
\label{appendix:coupler_dephase}
\begin{figure}[htp]
\includegraphics[width=8.6cm]{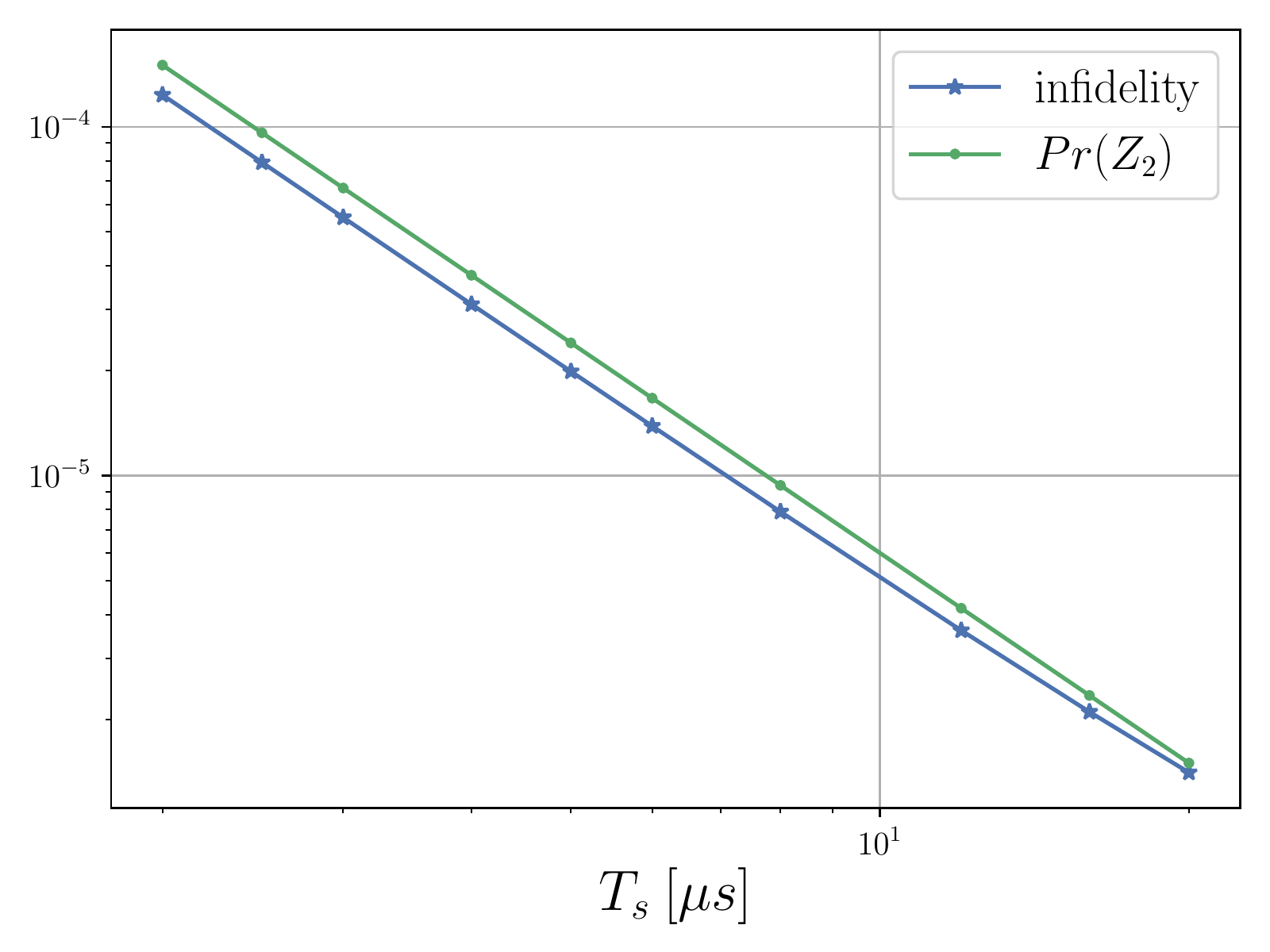}
\caption{The average infidelity and the Pauli channel probability $Pr(Z_2)$ (here qubit 2 is the ancilla) as a function of $T_s$ of the coupler (the coupler frequency is shifted by $2 \pi/T_s$), during the CZ gate that is presented in the main text. All the other Pauli channel probabilities for both qubits are smaller than $10^{-6}$ and are not shown.}
\label{fig: coupler_TS}
\end{figure}

In this section, we study the performance of the two-qubit gate (Sec.~\ref{subsec: 2q - Pulse shape and gate parameters}) under phase noise on the coupler, using the same method that we use to analyze noise in Fig.~\ref{fig:t_phi_and_threshold} in the main text.
As in the main text, we add a constant shift $\zeta c^\dagger c$, where $\zeta=\frac{2\pi}{T_s}$ and $c$ is the annihilation operator of the coupler, and show in Fig.~\ref{fig: coupler_TS} the average infidelity and the Pauli channel probability $Pr(Z_2)$  on qubit 2 (qubit 1 is the data qubit, while qubit 2 is the ancilla) under the Pauli-twirl approximation. All the other Pauli errors are below $10^{-6}$. As in the main text, the Twirled-Pauli channel probability is calculated by numerically building the process matrix ($\chi$) \cite{wood2015tensor_2}  of the gate (including the shift $\zeta c^\dagger c$), where $\chi$ is written in Pauli basis for qubits 1 and 2. Then, we extract the Twirled-Pauli channel probabilities by taking the diagonal values of $\chi$.
The parameters of the gate are the same as in Sec.~\ref{subsec: 2q - Pulse shape and gate parameters}. We see that the gate infidelity due to coupler dephasing is small ($\approx 10^{-4}$) and that most of the infidelity is explained by the dephasing of the ancilla ($Z_2$ error). The contribution to the overall infidelity is small because of the relatively small hybridization of the coupler with the other qubits (which is around $1\%$), see Fig.~\ref{fig:2q_gate_population_main}. The reason that the ancilla is affected more by the dephasing of the coupler is that the hybridization of the ancilla is stronger, since its frequency is closer to the coupler frequency [see also Fig.~\ref{fig:2q_gate_population_main}(b)]. Thus, since most of the infidelity is attributed to the dephasing of the ancilla, the surface code is less affected by it, as explained in the main text.

\section{calculating the threshold in a surface code}
\label{appendix: threshold}

In this section, we provide more details regarding the surface code threshold simulations we show in Fig.\ \ref{fig:t_phi_and_threshold}(b) in the main text. Our method to extract the threshold follows Ref.~\cite{PhysRevA.89.022321, PhysRevX.13.041022}. For the convenience of the reader, we reproduce here our noise model from the main text. We simulate the surface code using Stim \cite{gidney2021stim} and extract the threshold for three different models. In all of the models, we apply a depolarizing channel after every one-qubit operation and before every measurement, and a two-qubit depolarizing channel after every CZ, with the same error rate. For the ``$Z$-ancilla'' (``$Z$-data'') model, we also apply after each CZ a Pauli $Z$ with probability $Pr(Z)$ to the ancilla (data) qubit. In the ``readout'' model, before each ancilla measurement, we apply a Pauli $X$ with probability $p_{readout}=4Pr(Z)(1-Pr(Z))^3+4Pr(Z)^3(1-Pr(Z))$, to account for both cases of one or three $Z$ errors that lead to a measurement error. 

\begin{figure}[ht]
\centering
\includegraphics[width=8.6cm]{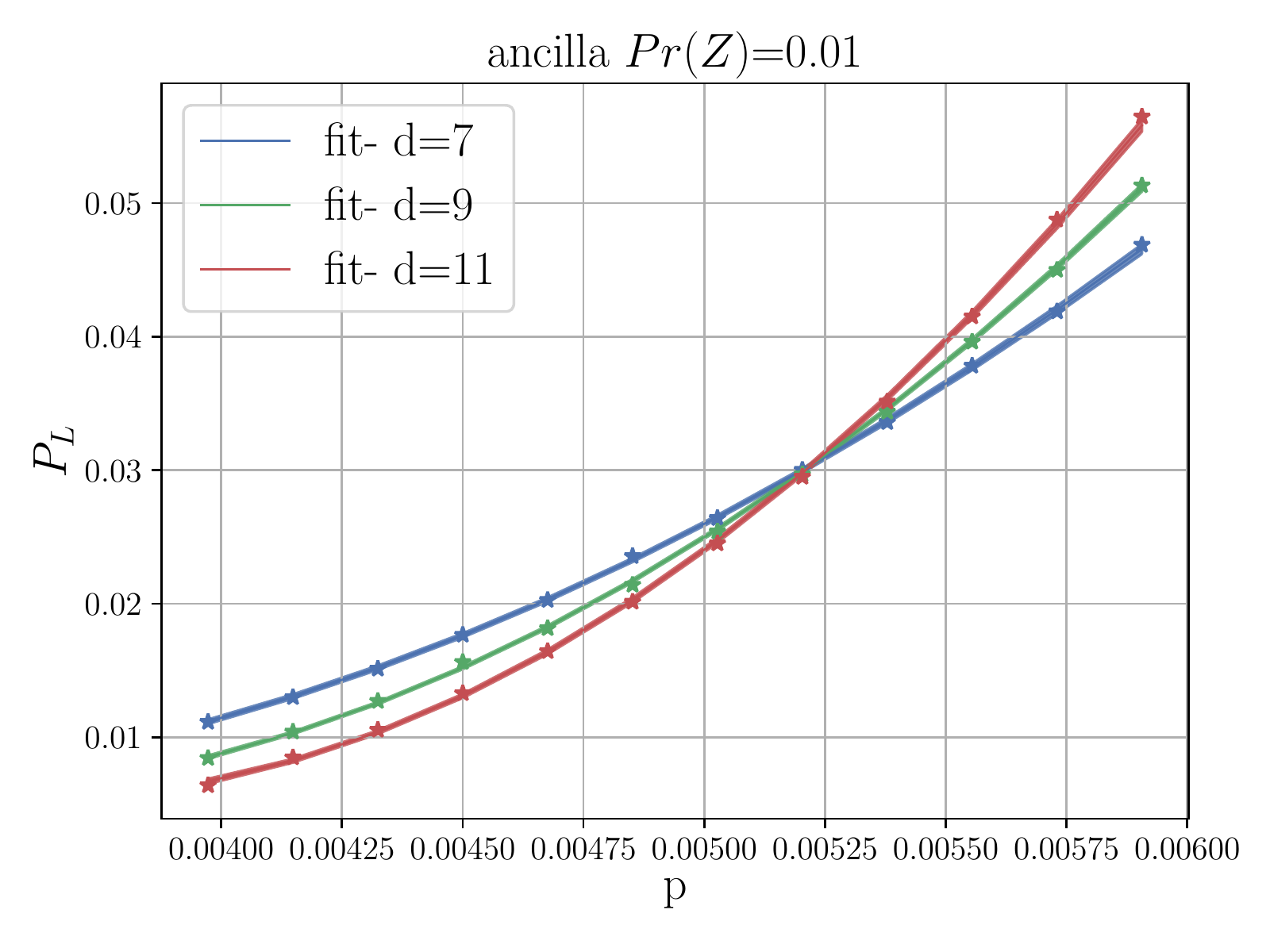}
\caption{An example for a simulation of the surface code for the "Z-ancilla" model, with $Pr(Z)=0.01$. The probability of the depolarizing channel is $p$, and the logical error probability is $P_L$. The stars are the values of the numerical simulation, and the 95\% confidence band of the fit is comparable to the thickness of each line. The parameters of the fit (which are independent of distance $d$) are $p_{th}=0.00523\pm 0.00001$ and $\mu=0.89\pm 0.02$.}
\label{fig:example_th}
\end{figure}

For each $Pr(Z)$, we extract the threshold $p_{th}$ as follows. As shown in Fig.~\ref{fig:example_th}, our data consists of the logical error probability $P_L$ at various distances $d$ and various probabilities of the depolarizing channel $p$. First, for every $d$ and every $p$, we compute a new scaled quantity $x=(p-p_{th})d^\mu$, where $p_{th}$ and $\mu$ are to be determined.

Then  we fit all data points (for this $Pr(Z)$) to
\begin{equation}
    P_L = A + Bx + Cx^2,
    \label{eq:threshold_calc}
\end{equation}
where $p_{th}, A, B, C, \mu$ are free parameters for the fitting and are independent of distance $d$. As in Ref.~\cite{PhysRevA.89.022321}, we use $d=7,9,11$ so as not to have small-distance effects. Figure~\ref{fig:example_th} shows the fit for $Pr(Z) = 0.01$. All of the points we present in Fig.\  \ref{fig:t_phi_and_threshold}(b) in the main text originated from a fit to Eq.~(\ref{eq:threshold_calc}). For each point in Fig.\ \ref{fig:t_phi_and_threshold}(b), the fit for each distance separately, using the same parameters for all distances (that were extracted using the scaled parameter $x$), gives $R^2\geq0.99$. 

\section{ramp-up and ramp-down pulses}
\label{appendix:ramp}
In this section, we provide more details regarding the ramp-up and ramp-down processes. We describe the pulses we use and show the evolution of state populations during the gate. 
In our simulations, we use the full cosine potential and thus use the Floquet states of the system as our eigenbasis. Therefore, the ramp-up and ramp-down pulses create a mapping between the Floquet states with the drive off (in which case the Floquet states are just the bare transmon states) and with the drive on (in which case the Floquet states are the spin-locked qubit states). We denote the instantaneous Floquet states (energies) for each point in time during the ramp pulse as $\{\ket{\hat{i}}\}$ ($\{\hat{E}_i\}$). These states are adiabatically connected to $\{\ket{{i}}\}$, the states of the transmon.  
  
As we describe in the main text, in order to prepare the correct computational Floquet states, we used Krotov's method \cite{GoerzSPP2019}  with the Hamiltonian in Eq.~(\ref{eq:full_cosine_Hamiltonian}) and the DRAG-inspired pulse as a starting point for the optimization. The control Hamiltonian is taken to be of the form
\begin{equation}
\left(-\Omega_{x}(t)\sin(\omega_{d}t)+\Omega_{y}(t)\cos(\omega_{d}t)\right)n.
\label{eq:krotov}
\end{equation}
We define $T_r$ as the ramp time. During ramp-up, $\Omega_{x}(0)=\Omega_{y}(0) = 0$, $\Omega_{x}(T_r)=\Omega$, and $\Omega_{y}(T_r)=0$. During ramp-down, the initial and final conditions are reversed. Here $\Omega$ and $\omega_d$ are the final values for our spin-locking continuous drive.

\begin{figure}[t]
\centering
\includegraphics[width=8.6 cm]{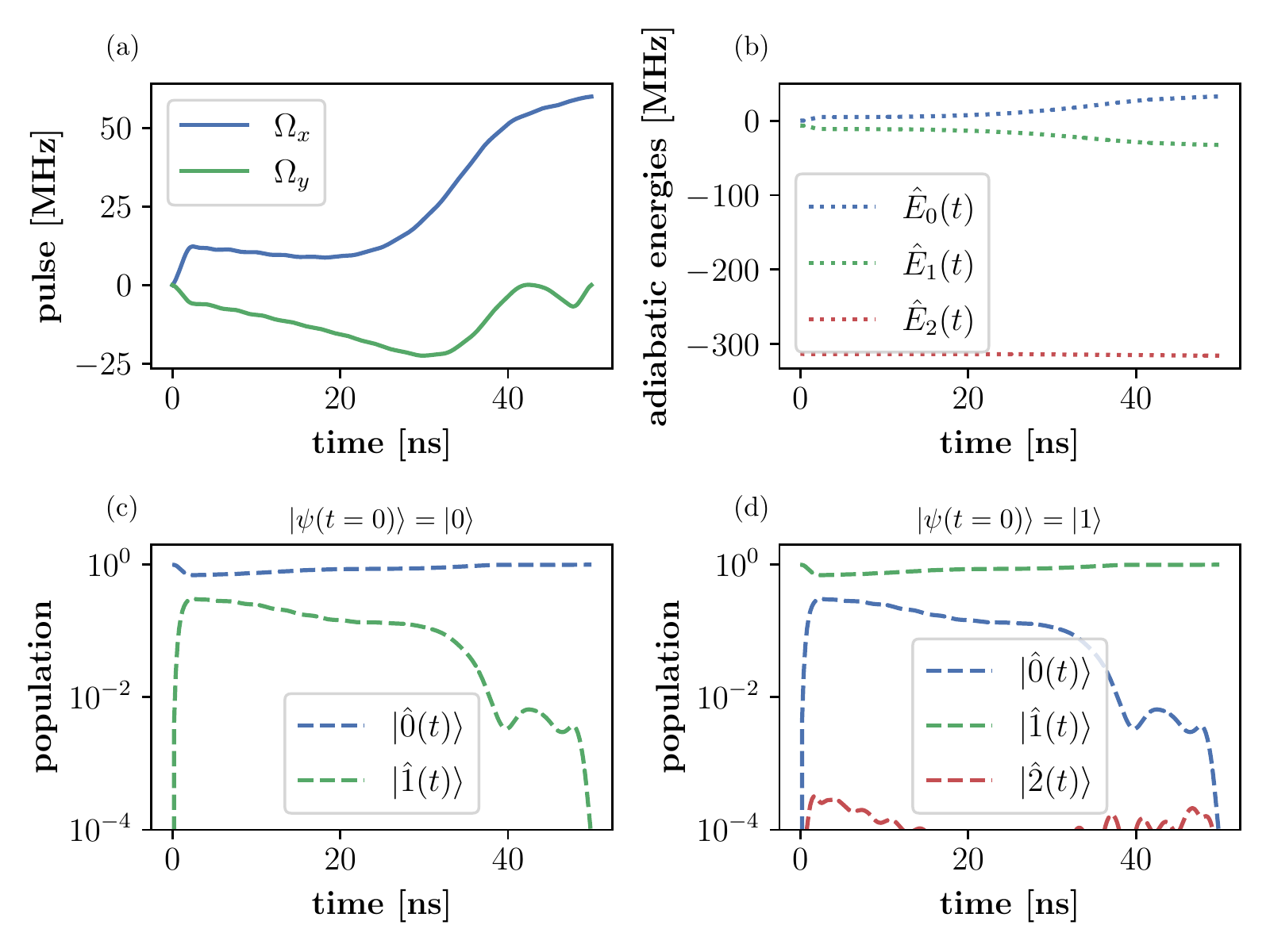}

\caption{Ramp-up pulse for $T_r=50$ ns. (a) The pulse envelopes $\Omega_x(t)$ and $\Omega_y(t)$, as defined in Eq.~(\ref{eq:krotov}). (b) The instantaneous Floquet energies. (c,d) The population of instantaneous Floquet states starting from (c) the ground state or (d) the first excited state of Eq.~(\ref{eq:full_cosine_Hamiltonian}). Although the points were evaluated only at integer multiples of $\frac{2\pi}{\omega_d}$, the graphs show smooth curves for readability.}
\label{fig:ramp_up}
\end{figure}

\begin{figure}[t]
\centering
\includegraphics[width=8.6 cm]{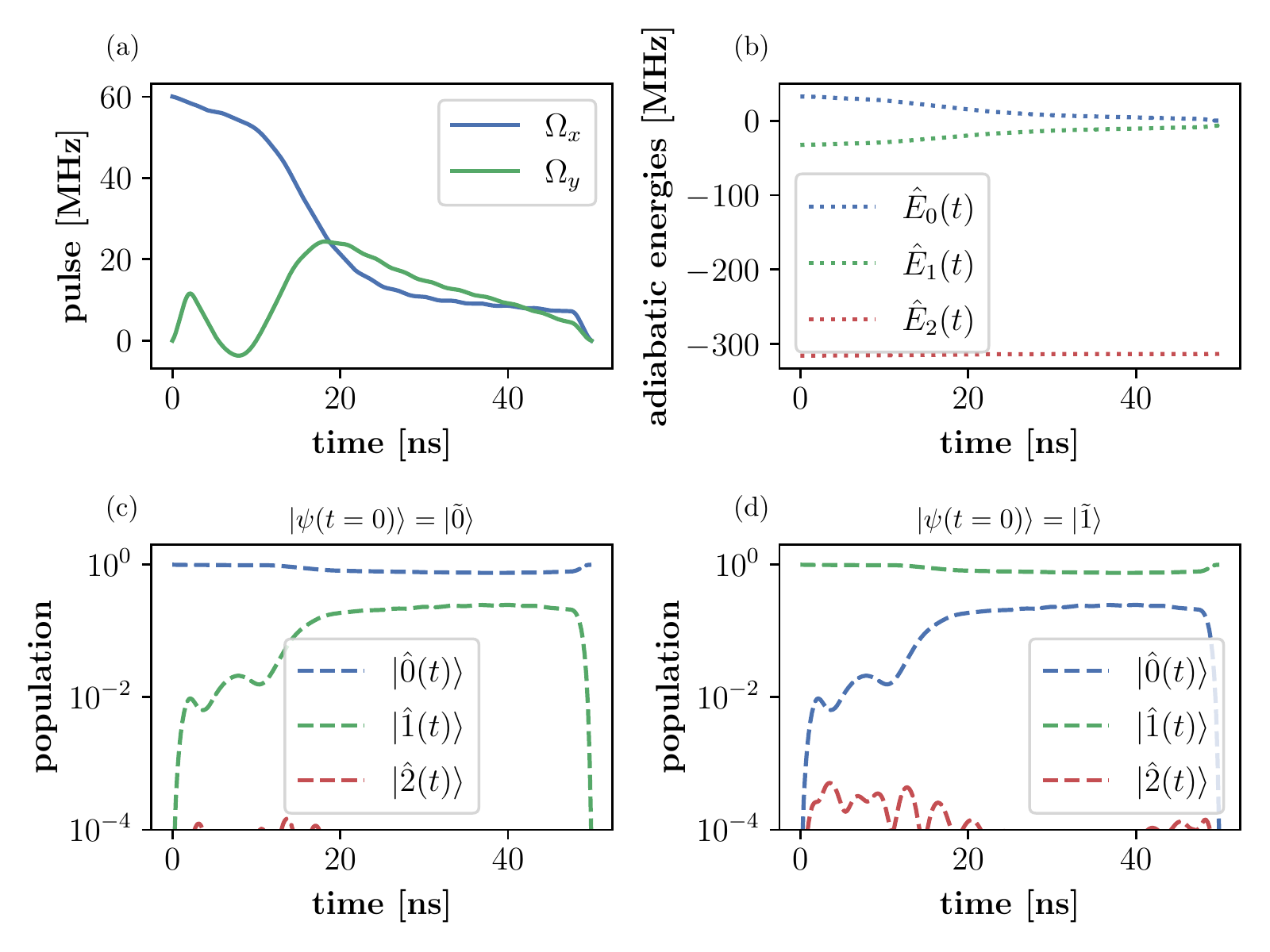}

\caption{Ramp-down pulse for $T_r=50$ ns. The figure is the same as in Fig.~\ref{fig:ramp_up}.}
\label{fig:ramp_down}
\end{figure}

Figures \ref{fig:ramp_up}(a) and \ref{fig:ramp_down}(a) show the pulses we found for ramp-up and ramp-down, with $T_r=50$ ns, respectively. We also plot in figures \ref{fig:ramp_up}(b) and \ref{fig:ramp_down}(b) the instantaneous Floquet energies, and in the corresponding (c) and (d) subfigures the instantaneous Floquet-state populations during the process. This means that, for each point in the graphs, we found Floquet energies and states that correspond to the current value of $\Omega_{x}$ and $\Omega_{y}$ in the pulse. We mark the instantaneous states and energies with a $\hat{}$ sign. So, during ramp-up, for example, if we start with $|0\rangle$, then $|\hat{0}(0)\rangle=|0\rangle$ and $|\hat{0}(T_r)\rangle=|\tilde{0}\rangle$ (recall that the $\tilde{}$ denotes the Floquet states of the driven Hamiltonian).    

Looking at the instantaneous populations and energies in Figs.~\ref{fig:ramp_up} and \ref{fig:ramp_down}, we can understand better what the optimized pulse sequence does. It starts by increasing the amplitude of the spin-locking drive $\Omega_x$ to increase the instantaneous energy gap. Then, it recovers the population using the DRAG-like pulse $\Omega_y$. 

\begin{figure}[t]
\centering
\includegraphics[width=8.6 cm]{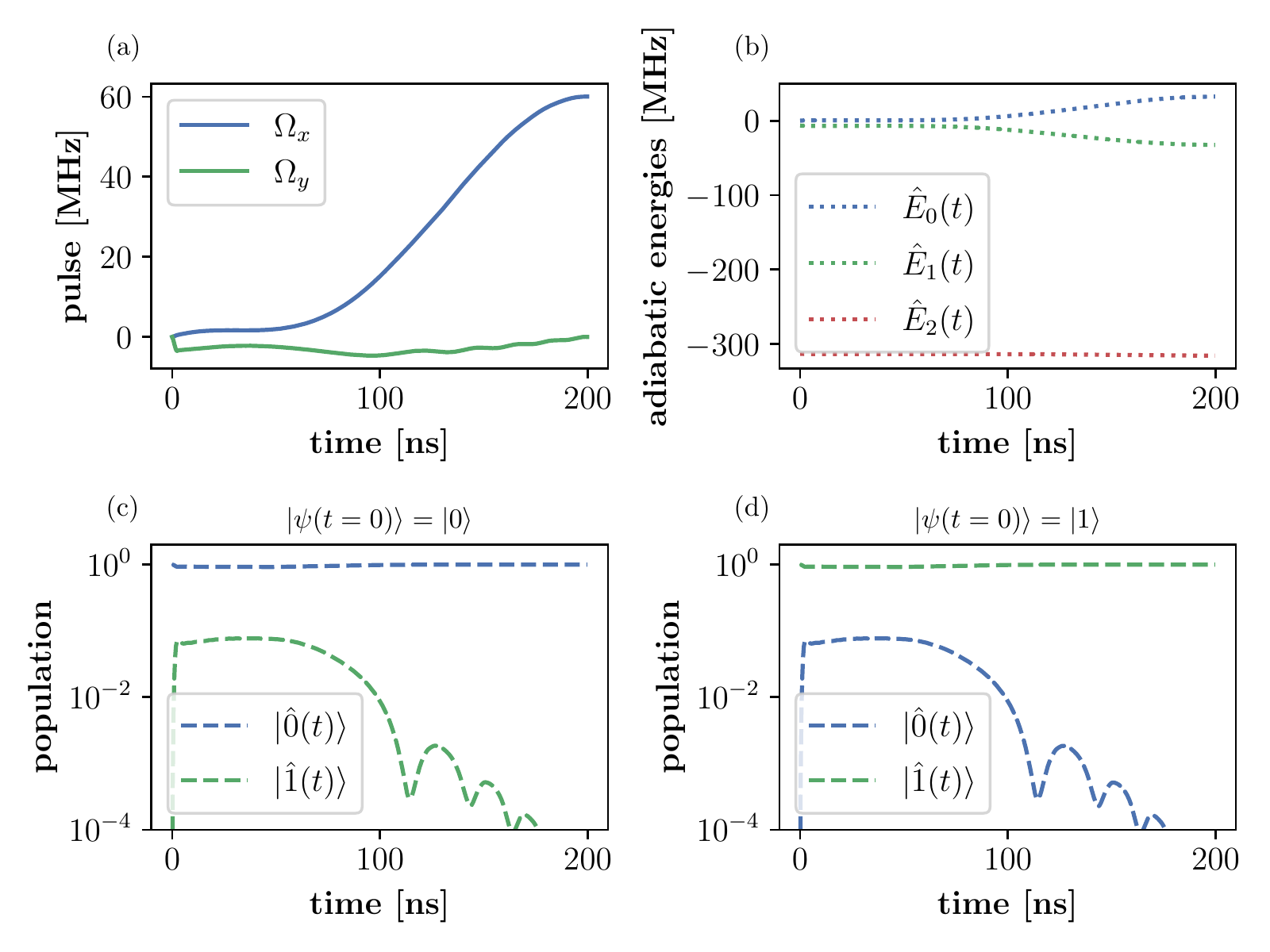}

\caption{Ramp-up pulse for $T_r=200$ ns. The figure is the same as in Fig.~\ref{fig:ramp_up}.}
\label{fig:ramp_up_slow}
\end{figure}

\begin{figure}[t]
\centering
\includegraphics[width=8.6 cm]{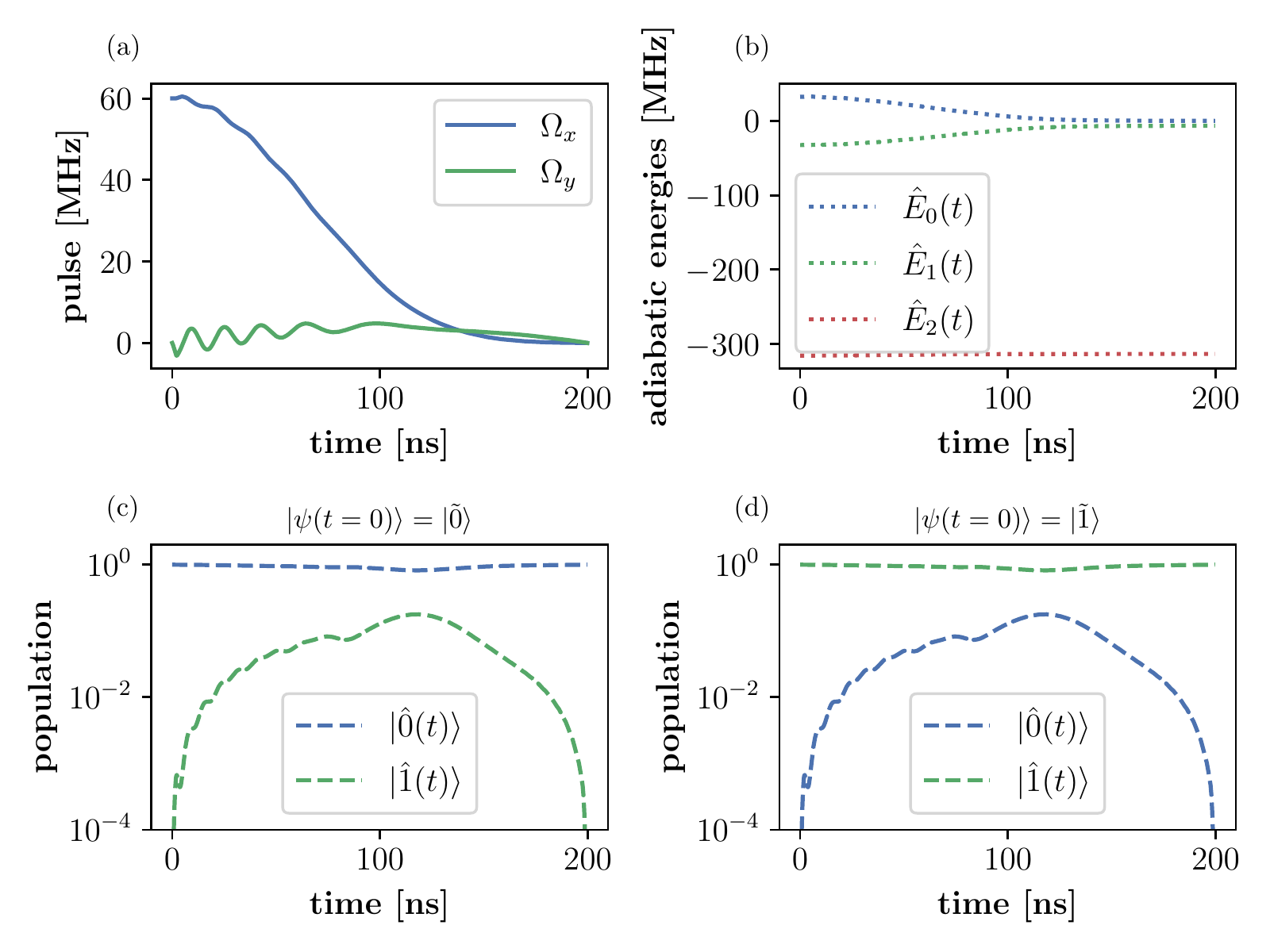}

\caption{Ramp-down pulse for $T_r=200$ ns. The figure is the same as in Fig.~\ref{fig:ramp_up}.}
\label{fig:ramp_down_slow}
\end{figure}

Additionally, Figures \ref{fig:ramp_up_slow} and \ref{fig:ramp_down_slow} show the same figures for $T_r=200$ ns. We see that the pulses are much smoother, as expected.

\section{Spin locking in the case of time-dependent transmon frequency}
\label{appendix:time-dependent transmon frequency}

In this section, we show that spin locking works seamlessly even if transmon frequency is time-dependent.

Consider a situation when the transmon frequency has known time dependence $\omega(t)$. Such a situation may arise, for example, if tuning a coupler alters the frequency of an adjacent transmon. To spin-lock such a transmon, we drive it as follows:
\begin{equation}
H=\omega(t) a^{\dagger}a-\frac{\eta}{2}a^{\dagger2}a^{2}+\frac{\Omega}{2}(ae^{i\phi(t)}+h.c),
\end{equation}
where the phase $\phi(t)$ is chosen such that $\dot \phi(t) - \omega(t) = \delta(\Omega)$ at all $t$, where $\delta(\Omega)$ satisfied the clock condition in Eq.\ (\ref{eq:insensativity}) in the main text.

To analyze this Hamiltonian, we go into an interaction picture defined by $U = e^{i \phi(t) a^\dagger a}$, so that the interaction-picture state is $\ket{\tilde \psi} = U \ket{\psi}$ and the interaction-picture Hamiltonian is
\begin{eqnarray}
 \tilde H &=& i \dot U U^\dagger + U H U^\dagger \\
 &=& (\omega(t)-\dot \phi(t)) a^{\dagger}a-\frac{\eta}{2}a^{\dagger2}a^{2}+\frac{\Omega}{2}(a+h.c) \\
 &=& -\delta a^{\dagger}a-\frac{\eta}{2}a^{\dagger2}a^{2}+\frac{\Omega}{2}(a+h.c),
\end{eqnarray}
which is exactly Eq.\ (\ref{eq:RWA_Hamiltonian}) in the main text.

This means that, as long as we precisely know the time-dependent frequency $\omega(t)$ of the transmon and can precisely engineer the drive to have the desired phase $\phi(t)$, spin locking works seamlessly even if transmon frequency is time-dependent.  

\section{Measurement of the dressed qubit}
\label{appendix:measure}

In the main text, we assume that the dressed qubit is measured by first ramping down the dressing fields. In this section, we show how to do the measurement of the dressed qubit without turning off the dressing.

Suppose that our spin-locked transmon is coupled to a one-sided cavity. The Hamiltonian of the system in the Kerr approximation and in the rotating-wave approximation is
\begin{equation}
H=\omega a^{\dagger}a-\frac{\eta}{2}a^{\dagger2}a^{2}+\frac{\Omega}{2}(ae^{i\omega_dt}+h.c.) + \omega_c b^\dagger b + g (a^\dagger b + h.c.),
\end{equation}
where $a$ is the annihilation operator for the transmon (as in the main text) and $b$ is the annihilation operator for a cavity of frequency $\omega_c$. Here $g$ is the coupling between the transmon and the cavity. The Heisenberg-Langevin equation of motion for $b$ is
\begin{equation}
\dot b = - \frac{\kappa}{2} b - \sqrt{\kappa} b_\textrm{in} + i [H,b],
\end{equation}
where $b_\textrm{in}$ is the input field and $\kappa$ is the cavity linewidth. The input-output relation expresses the output field $b_\textrm{out}$ in terms of the input field and the cavity field:
\begin{equation}
b_\textrm{out} = b_\textrm{in} + \sqrt{\kappa} b.
\end{equation}

Moving, as in the main text, into a frame rotating with frequency $\omega_d$, we get
\begin{align}
   H_I =- \delta a^{\dagger}a-\frac{\eta}{2}a^{\dagger2}a^{2}+\frac{\Omega}{2}(a+a^\dagger) - \delta' b^\dagger b + g (a^\dagger b + a b^\dagger), 
\end{align}
where $\delta = \omega_d - \omega$ and $\delta' = \omega_d - \omega_c$.

Let's assume that $\kappa \ll g \ll \Omega$. We then start by diagonalizing the dressed transmon to obtain dressed eigenstates $\ket{\tilde i}$ and eigenenergies $\tilde E_i$ with $i = 0, 1, 2, \dots$, where $i = 0,1$ correspond to the dressed qubit insensitive to phase fluctuations to first order. Recall that the energies are sorted from the highest to the lowest. The Hamiltonian can therefore be written as 
\begin{align}
   H_I = \sum_i \tilde E_i \ket{\tilde i} \bra{\tilde i} - \delta' b^\dagger b + g (a^\dagger b + h.c.), 
\end{align}
where the original transmon annihilation operator $a$, when written in the dressed basis, has all entries being generically nonzero.

Let's now choose the cavity frequency $\omega_c$ such that $\delta' = \tilde E_1 - \tilde E_2$, i.e.\ we assume the cavity is resonant with the $\ket{\tilde 1} \rightarrow \ket{\tilde 2}$ transition. Let's now send into the cavity (via $b_\textrm{in}$)  weak light resonant with the cavity. Let's further assume that $g \ll \tilde E_1 - \tilde E_2$ and that all other transitions involving dressed states $\ket{\tilde 0}$ and $\ket{\tilde 1}$ (i.e.\ $\tilde E_0 - \tilde E_{j > 0}$ and $\tilde E_1 - \tilde E_{j > 2}$) differ in frequency from $\tilde E_1 - \tilde E_2$ by an amount much larger than $g$. This allows us to keep in the Hamiltonian only one resonant term generated by $a^\dagger$:
\begin{align}  
   H_I = \sum_i \tilde E_i \ket{\tilde i} \bra{\tilde i} - \delta' b^\dagger b + (\tilde g \ket{\tilde 2} \bra{\tilde 1}  b + h.c.), 
\end{align}
where $\tilde g = g \bra{\tilde 2} a^\dagger \ket{\tilde 1}$.

If the dressed transmon is in state $\ket{\tilde 0}$, then the cavity essentially doesn't see the transmon (because we assumed all transitions involving $\ket{\tilde 0}$ are far off-resonance from the cavity). Therefore, incoming light $b_\textrm{in}$ scatters essentially from an empty cavity. Changing the frequency of the rotating frame from $\omega_d$ to $\omega_c$ and treating $b$, $b_\textrm{in}$, and $b_\textrm{out}$ as complex numbers describing classical coherent states, the resulting equations are
\begin{eqnarray}
    \dot b &=& - \frac{\kappa}{2} b - \sqrt{\kappa} b_\textrm{in},\\
    b_\textrm{out} &=& b_\textrm{in} + \sqrt{\kappa} b.
\end{eqnarray}
Solving the first equation in steady state for $b$ and plugging into the second equation, we find
\begin{align}
b_\textrm{out} = -b_\textrm{in},
\end{align}
which shows that the light picks up a phase of $\pi$ by reflecting off an empty resonant cavity.

Now suppose the dressed transmon is in state $\ket{\tilde 1}$. Since we assumed that the $\ket{\tilde 0}$-$\ket{\tilde 1}$ transition is far off-resonance from the cavity, the presence of the cavity in the vacuum state $\ket{0}$ doesn't affect the dressed qubit. On the other hand, when we probe the cavity with weak light, we need to consider the Hilbert space associated with the one-photon Fock state $\ket{1}$ of the cavity. Since we assumed $g \gg \kappa$ and since $\bra{\tilde 2} a^\dagger \ket{\tilde 1}$ is of order 1, we have $\tilde g \gg \kappa$ and should therefore first diagonalize the $\tilde g$ interaction between the dressed transmon and the cavity. The coupling $\tilde g$ resonantly couples $\ket{\tilde 1} \ket{1} $ to $\ket{\tilde 2} \ket{0}$ resulting in eigenstates  $\ket{\tilde 1} \ket{1} \pm \ket{\tilde 2} \ket{0}$ shifted by $\pm \tilde g$. Since $\tilde g \gg \kappa$, the light that is resonant with an empty cavity can thus no longer enter the cavity. Treating again $b$, $b_\textrm{in}$, and $b_\textrm{out}$ as complex numbers describing classical coherent states, we therefore have $b = 0$, which in turn means 
\begin{equation}
b_\textrm{out} = b_\textrm{in},
\end{equation}
which shows that the light doesn't pick up the $\pi$ phase if it scatters off an off-resonant cavity. 

By detecting the phase of the scattered light, we can therefore detect whether the dressed transmon is in state $\ket{\tilde 0}$ (which gives a $\pi$ phase shift) or in state $\ket{\tilde 1}$ (which gives no phase shift).

\section{4-body terms due to ZZ cross-talk in a non-commuting gate realization}
\label{appendix:ZZ cross-talk}
\begin{figure}[ht]
\includegraphics[width=8.6cm]{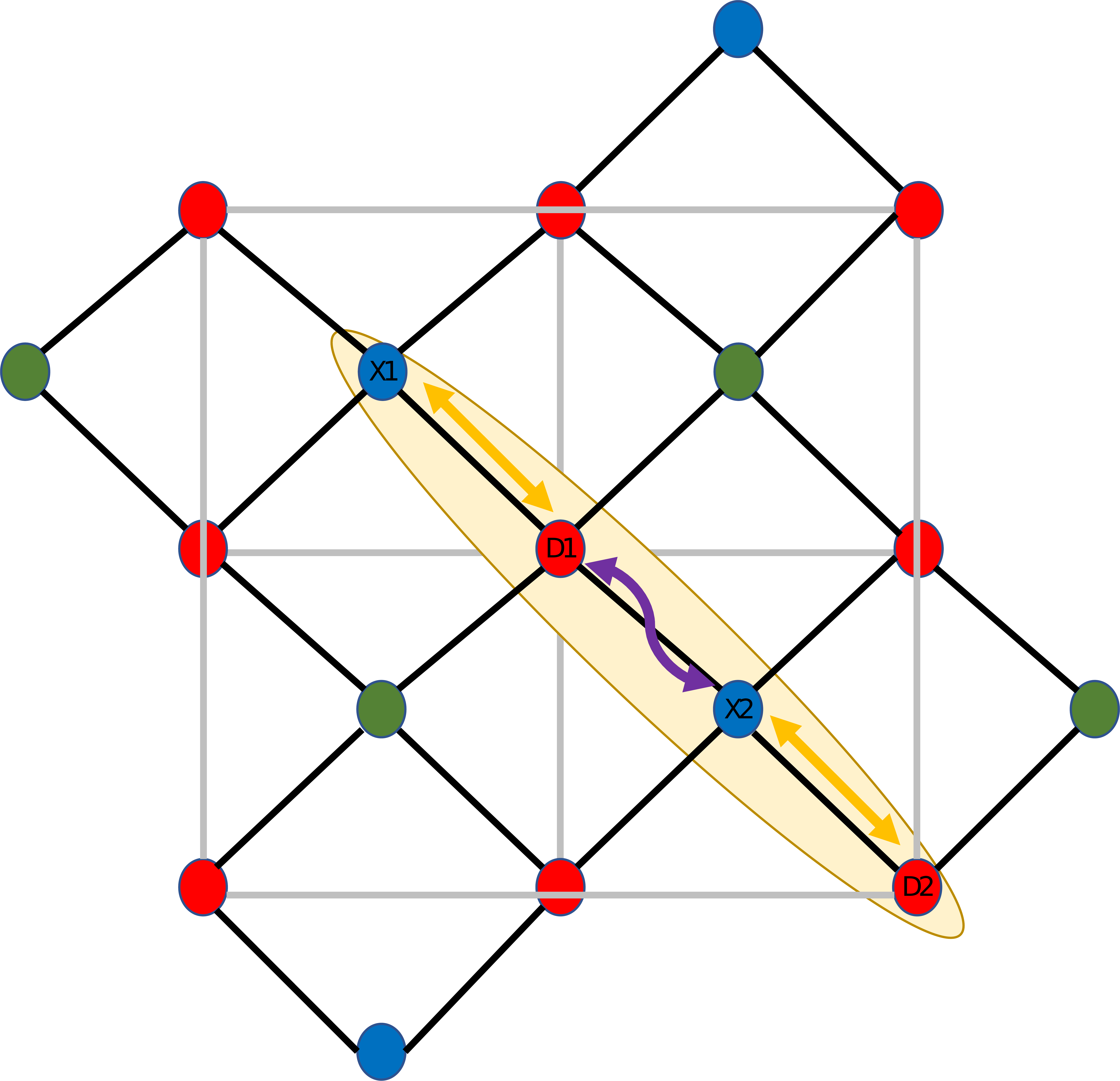}
\caption{An example for a distance-3 surface code. The red circles are the data qubits, and the blue (green) are the ancilla qubits associated with the X (Z) stabilizers. The relevant qubits are in the bright yellow area. The orange arrows represent a gate between the qubits, while the curved purple arrow represents the ZZ cross-talk.}
\label{fig: surface_code}
\end{figure}

In this section, we will explain how 4-body terms (i.e., Pauli strings with weight four) could arise in a surface code context due to ZZ cross-talk, and why the adiabatic ZZ gate is robust to such errors. In this discussion, we assume there is some residual ZZ cross-talk between any two neighboring qubits. Although this cross-talk has already been studied numerically \cite{PRXQuantum.3.020301}, here we show a qualitative simplified analysis that concentrates on calculating the error probabilities under the Pauli-twirl approximation. The structure of the surface code enables performing some of the gates simultaneously, thus reducing time overhead, as can be seen in Ref.~\cite{PhysRevApplied.8.034021}. As a consequence, there are "chains" of gates that operate simultaneously and reside on the diagonals of the code, as can be seen in the bright yellow area in Fig.~\ref{fig: surface_code}, which shows an example of a distance-3 rotated surface code \cite{PhysRevA.90.062320}. Due to ZZ cross-talk in the system, the links in the chains are connected through the residual interaction. Fig.~\ref{fig: surface_code} shows such an example chain. The ancilla qubit X1 is connected via a gate (orange arrow) with data qubit D1, which is in turn  connected via a ZZ interaction (curved purple arrow) with ancilla qubit X2, which is in turn connected via a gate (orange arrow) with data qubit D2. We will limit our discussion only to chains of four qubits, for reasons that will be clear below.

Assume that the gate interaction is a Hermitian operator $G$, the gate strength is $\xi$, and the ZZ strength is $g_{zz}$. Assume that we want to realize a CZ gate, so that the gate unitary is $e^{-it\xi G}=$ CZ for some predetermined time $t$. We can model the chain Hamiltonian as
\begin{equation}
    H = \xi G^{X1,D1}+g_{zz}\sigma_z^{D1}\sigma_z^{X2}+\xi G^{X2,D2},
\end{equation}
where $G^{X1,D1}$ is the gate interaction between X1 and D1 (similarly for X2 and D2). Therefore, the time evolution of the system is
\begin{equation}
    U(r) = e^{-itH}=e^{-i\theta(G^{X1,D1}+G^{X2,D2}+r\sigma_z^{D1}\sigma_z^{X2})},
\end{equation}
where we introduced $r=\frac{g_{zz}}{\xi}$ and $\theta=\xi t$ being the rotation angle of the gate. Here $r=0$ would correspond to two ideal CZ gates.  For example, in the adiabatic ZZ gate with $G = ZZ$, the rotation angle is $\theta=\frac{\pi}{4}$. On the other hand, in a gate that utilizes a full rotation around the $|02\rangle$ state using $G =|11\rangle\langle02| + h.c.$, the rotation angle is  $\theta=\pi$. Assuming $r\ll1$, we wish to understand the error channel that arises due to the cross-talk. Thus, to factor out the desired dynamics, we multiply $U(r)$ by $U^{-1}(r=0)$, where $U(r=0)=U_0$ is the ideal unitary. Thus,
\begin{equation}
    U_0^{-1}U = e^{i\theta(G^{X1,D1}+G^{X2,D2})} e^{-i\theta(G^{X1,D1}+G^{X2,D2}+r\sigma_z^{D1}\sigma_z^{X2})}. \label{eq:U0U}
\end{equation}
Assuming we could expand $U_0^{-1}U$ using a Taylor expansion in $r$, and denoting $H_0=G^{X1,D1}+G^{X2,D2}$, we get
\begin{eqnarray}
    U_0^{-1}U &\approx& 1-i\theta r\int_0^1 e^{i\theta (1-\alpha)H_0} \sigma_z^{D1}\sigma_z^{X2} e^{-i\theta (1-\alpha)H_0}d\alpha \nonumber \\ 
     &=& 1-i\theta r\int_0^1 \sum_{m=0}^\infty \frac{(i\theta (1-\alpha))^m}{m!} L_{H_0}^m(\sigma_z^{D1}\sigma_z^{X2}) d\alpha, \nonumber \\& \label{eq:taylor_U}
\end{eqnarray}
where $L_{H_0}^0(A) = A$ and $L_{H_0}^m(A)=[H_0,L_{H_0}^{(m-1)}(A)]$ for any operator $A$ and $m\geq1$ \cite{Casas_2012_revised}, and the integral comes from the derivative of the exponential map \cite{Wilcox:1967zz}. To calculate the resulting Pauli error channel under the Pauli-twirl approximation, we need to calculate
\begin{equation}
    p_{ijkl}= \left|\frac{1}{2^4}Tr(\sigma_i^{X_1}\sigma_j^{D_1}\sigma_k^{X_2}\sigma_l^{D_2}U_0^{-1}U)\right|^2,
    \label{eq: prob_PTA}
\end{equation}
where $p_{ijkl}$ is the probability for a $\sigma_i^{X_1}\sigma_j^{D_1}\sigma_k^{X_2}\sigma_l^{D_2}$ error under the Pauli-twirl approximation, where $\sigma_m \in \{I, \sigma_x, \sigma_y, \sigma_z\}$.
 If we want to consider only errors that have a  probability of order $r^2$ (which is the first non-vanishing order), it is clear from Eq.~(\ref{eq:taylor_U}) and Eq.~(\ref{eq: prob_PTA}) why we need to take into account in this model no more than 4 qubits. Any interaction that involves more than 4 qubits under this model results in using two different ZZ links (two purple arrows in Fig.~\ref{fig: surface_code}), and thus would result in error probabilities that are $\mathcal{O}(r^3)$. From Eq.~(\ref{eq:U0U}), it can be seen that, if $[H_0, \sigma_z^{D1}\sigma_z^{X2}]=0$, then $U_0^{-1}U=e^{-ir\theta \sigma_z^{D1}\sigma_z^{X2}}$. Thus, the only non-trivial error 
would be $p_{IZZI}$, and the maximal weight of any erroneous Pauli string under this analysis throughout the full surface code would be two (up to order $r^2$, since 4 qubits may not be sufficient at higher order). On the other hand, if $[H_0, \sigma_z^{D1}\sigma_z^{X2}]\neq0$, that would not be necessarily true, and also 3- or 4-body error probabilities could arise up to order $r^2$, as can be seen from Eq.~(\ref{eq:taylor_U}). 

As an example, assume $G=|11\rangle\langle02| +h.c$ and $\theta=\pi$, i.e., the gate is a full rotation around the $|02\rangle$ state to create a CZ gate. Thus, using Eq.~(\ref{eq: prob_PTA}), we get, up to order $r^2$,
\begin{equation}
    p_{ZZZZ} = \frac{9\pi^2}{4096}r^2.
\end{equation}
Moreover, the total amount of errors is $1-p_{IIII}=\frac{2519\pi^2}{4096}r^2$, but $p_{IZZI}=\frac{1849\pi^2}{4096}r^2$. That means that approximately 25\% of the error probability is distributed on different non-trivial Pauli strings with weights up to 4. 
This example shows that gates that commute with the cross-talk operator would limit the creation of multi-body error terms, which could be harmful, especially for low-distance codes, where two of these 4-body Pauli strings are supported on data qubits. As future research, a more thorough and numerical analysis could be made to understand the impact of these errors on the code. We note that the adiabatic ZZ gate commutes with the noise operator only in the effective Hamiltonian picture,  
while the full model would still give rise to such multi-body errors, but those are expected to be smaller than in the case where the native gate does not commute with ZZ 
in the effective picture.


\bibliography{ms}


\end{document}